# Handbook on Best Practice for Minimising Beam Induced Damage during IBA

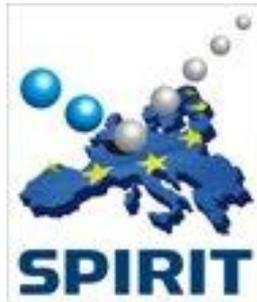

*Editors*
*D. Benzeggouta, I. Vickridge*
*Université de Pierre et Marie Curie,*
*UMR7588 du CNRS, Paris*

*Contributors*
*N. P. Barradas*
*D. Benzeggouta*
*M. Doebeli*
*C. Jeynes*
*A. Vantomme*
*I. Vickridge*

# *Contents*

# 1. Introduction

"Ion Beam Analysis" encompasses a suite of analytical techniques that make use of the interactions between beams of rapid ions and a sample to be analysed. There is no definitive list or even perimeter of what is universally accepted to fall within the fold of IBA, however here, we consider it to be restricted to prompt interactions, with beams that can be produced by electroststatic accelerators. From an operational point of view, this includes Particle-Induced X-ray Emission (PIXE), Particle-Induced gamma-ray Emission (PIGE), Rutherford Backscattering Spectroscopy (RBS), Nuclear Reaction Analysis (NRA) and the various classes of Elastic Recoil Detection Analysis (ERDA). Typically ion beams of energy between 0.1 and 100 MeV are used. By this definition we specifically exclude Low Energy Ion Scattering (LEIS) and Secondary Ion Mass Spectroscopy (SIMS). To this list of exclusions we also add Accelerator Mass Spectrometry (AMS), which is a destructive IBA method.

When an ion beam analyst interrogates a sample with a charged particle beam, only a tiny fraction of the incident energy is harvested in terms of analytical information. For example, when performing PIXE on silicon with 1 MeV protons, about 1 proton in 10 generates a k-shell ionization in a silicon atom. Only about 1 in 20 of these vacancies is filled by a radiative process leading to characteristic x-ray emission, and a typical solid state x-ray detector will subtend at most about $1/10^{th}$ of the overall $4\pi$ solid angle into which the x-rays are emitted. Thus, to observe 1 Si k x-ray of about 2 keV, it is necessary to send in 2000 protons of 1 MeV – i.e. to dissipate about 1 million times more energy than is actually collected for the analysis. The serious ion beam analyst generally masters the interactions that lead to the analytical signal (x-rays, elastically scattered particles, nuclear reaction products etc) but also needs to be aware of the other consequences, for the sample under analysis, of the dissipation of the incident energy. In Ion Beam Analysis, a strict view of the situation would term the sum of the consequences of the impingent energy 'sample damage'. In this strict view, an ion beam analyst inflicts vastly more damage on a sample than he recovers information. This should never be ignored.

Taking a severe stance on what we consider to be damage, makes it incumbent on us to consider the nature and importance of the damage in *any* given analysis. Ion beam analysts at first described IBA as 'non-destructive', so as to differentiate it from methods such as wet analytical chemistry, pyrolysis, or sputtering methods that needed to consume the sample in order to analyse it. This is a good seller's line, but the term was sometimes interpreted to mean 'no damage' at which point it is no longer clear what each analyst might mean by 'non-destructive', nor that the analyst and the owner of the sample being analysed can agree whether or not the sample has been 'damaged' by the analysis. Does it mean that there is no loss of atoms from the sample? Does it mean that the analysis will give the same results when repeated once? twice? indefinitely? Does it mean that the owner of the sample cannot detect that the sample has been analysed? What about latent damage that is revealed only at some time after the analysis?

Of course, in any useful analysis, the value of the analytical results to be obtained is weighed against the nature and significance of the damage that will be incurred; the ion beam analysis techniques exist because there are many situations where the analytical value far outweighs the cost of the damage. However, even in these cases, it is usually desirable to limit the damage for a given analysis.

It is the purpose of this handbook, to give the ion beam analyst some basic tools and information, that will allow a fair assessment of the damage likely to be incurred during an analysis, together with a sufficient understanding of the underlying processes that best damage mitigation practices may be developed and employed.

From an operational point of view, we divide the damage into two types:

We consider to be **Type I**, damage that influences an IBA measurement, and thus may be observed by its effects on the measurement, during the measurement. Examples are crystal damage incurred during channeling measurements, sputtering and other element loss, radiation-induced diffusion during the measurement, and sample deformation such as shrinking and buckling typical of biological materials.

We consider to be **Type II**, damage that occurs in the sample due to the IBA measurement, that is not discernible by its effect on the measurement. Examples are changes in optical, magnetic or electrical properties, micro-structural changes such as polymerization or densification in amorphous materials, changes in the chemical states of atoms and the bonds that tie them together, and loss or incorporation of atomic species (e.g. H) that are not visible with the IBA method being used. Type II damage may be latent – such as in the case of chemical changes that lead to reactivity of the sample with air after analysis in vacuum or under inert gas.

A special case is carbon deposition under the beam. Except in the rare cases where IBA is performed under well-controlled ultra-high vacuum conditions (for example a residual pressure of $10^{-10}$ mbar composed largely of $H_2$ and CO) the residual gas in the target chamber ($10^{-6}$ to $10^{-8}$ mbar) often contains significant partial pressures of organic carbon-based molecules – mostly oils from pumps or organic molecules transferred from hands when handling target chamber components without using gloves. These molecules may be cracked by the beam and deposited on the sample being analysed.

In some cases the damage may be partially or completely recovered after the measurement – for example by thermal annealing in the case of damage to a crystal and in some cases of optical and electrical damage.

This document is organized into two main sections. The first section covers the basic physical processes that occur during conversion of the beam kinetic energy into, ultimately, thermal energy dissipated in the sample and the surrounding environment, and potential energy stored in the analysed sample. This is followed by a brief overview of damage from the point of view of various IBA methods. The second section discusses the specificities of damage induced by IBA of various classes of material – polymers, single crystals, metals, semiconductors and so on, together with examples and specific suggestions to help in maximizing the 'information to damage ratio'.

A small but significant third section is devoted to the important question of uncertainty in IBA data. This is deeply tied to the question of damage incurred by analysis of the sample. One may pose the question this way: what is the minimum beam dose (and therefore damage) that is required to provide an answer to my analytical question, to within a specified level of uncertainty? Contained in this question are the questions "How much can I be sure that this spectrum is telling me?" and "How much can I be sure this spectrum is NOT telling me?". It is not suggested that every IBA analysis need be subjected to such Deep Thought, however the analyst needs to know when and how to pose these questions, and how to obtain reasonable answers to them.

### 1.1 Overview of basic damage production mechanisms during IBA

There are many reference texts concerning ion beam damage and ion beam energy dissipation processes in matter. The standard IBA references such as (Chu et al. 1978; Wang and Nastasti 2010) unfailingly deal with stopping power. A dated but still very useful reference on defect production is (Lehman 1977), which covers the many facets of the physics of energy loss and defect production at a comfortable level of sophistication for physics undergraduates. Published in 1977, just prior to the explosion of the use of Monte Carlo calculations, it provides good physical insight into the underlying processes. In the following sections, it is not our aim to reproduce what is already available in such texts, but rather to provide a succinct overview that leaves the reader with the background necessary to understand the subsequent sections based on damage production in different materials, and to identify for themselves areas where further reading will be helpful in their specific analytical context.

### Setting the scene

An ion loses its energy ($E$) nearly continuously along its path $x$, through electromagnetic interactions with the solid, considered here as constituted of a collection of randomly distributed atoms. The average energy loss per unit path length is termed the ''stopping power'' $S$ of the material for the ion. A semi-empirical synthesis of experimental values for all ions in all elements is available (SRIM) and a simple treatment is given in (Vickridge and Schmaus 2006). Briefly, $S$ may be deduced from the measured stopping cross-section:

$$\varepsilon = -\frac{1}{N}\frac{dE}{dx} \quad \text{(eV.cm}^2\text{)} \tag{1}$$

with $N$ the atomic density of the solid: $S = -N\varepsilon = -dE/dx$ (eV/nm). There are two distinct and independent contributions to $S$: interactions with the electrons of the materiel, contributing what is called the electronic stopping power $S_e$, and interactions with the atomic nuclei of the material contributing the nuclear stopping power $S_n$ so that $S = S_e + S_n$. $S_e$ leads to ionization within the target, and includes losses to collective electron excitations – plasmons. In a view in which the incident ion is considered to suffer a series of binary collisions with isolated atoms, $S_e$ is sometimes called inelastic energy loss in the literature, since the ion-atom collision is no longer elastic, involving energy exchanges with the electronic system of the atom. By the same token, in this view $S_n$ is called elastic energy loss.

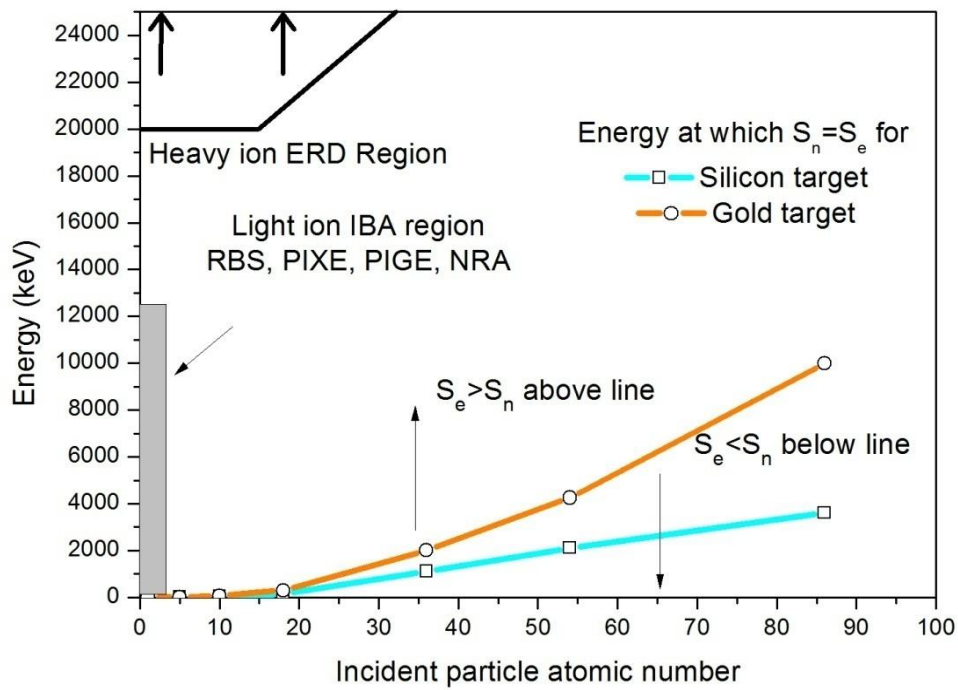

***Figure 1 :*** Energy-atomic number plot showing various IBA domains.

The two contributions dominate the energy loss in different regions of ion-target combinations and charged particle velocity, and have quite different damage-inducing mechanisms and effectiveness.

Figure 1 traces the energy at which $S_e = S_n$ for ions incident on a light (Si) and heavy (Au) target, as a function of incident ion atomic number, calculated with SRIM. The further from the line (above, or below) the greater is the difference between $S_e$ and $S_n$ Also shown are the regions where IBA is typically performed. It is clear that for practically all IBA situations $S_e$ largely dominates $S_n$.

To these two major contributions to $S$, we may add the very small contribution from nuclear reactions, which may cause transmutation of nuclei and radio-activation of samples. Although representing only an infinitesimal fraction of kinetic energy deposited into the samples, significant further energy is contributed locally by the energy liberated by the nuclear reaction. The consequences of these effects when present can be major: activated samples require stringent handling procedures under standard radioprotection regimes.

In order to consider the order of magnitude of the time, energy and space scales involved, consider an alpha particle travelling in silicon. In Figure 2 we plot its stopping power and range as a function of its kinetic energy. A 2 MeV alpha particle will deposit all of its energy along a track about 7 μm long. Its velocity (at 2 MeV) is about $10^7$ m/s, or 0.033c. At the start, it loses energy at a rate of around 25 eV/nm, rising to a maximum rate of around 34 ev/nm when it has slowed down to around 500 keV, after which the rate of energy loss decreases down to zero. Practically all of this energy loss is electronic. The entire process takes a time on the order of $2 \times 10^{-12}$ s.

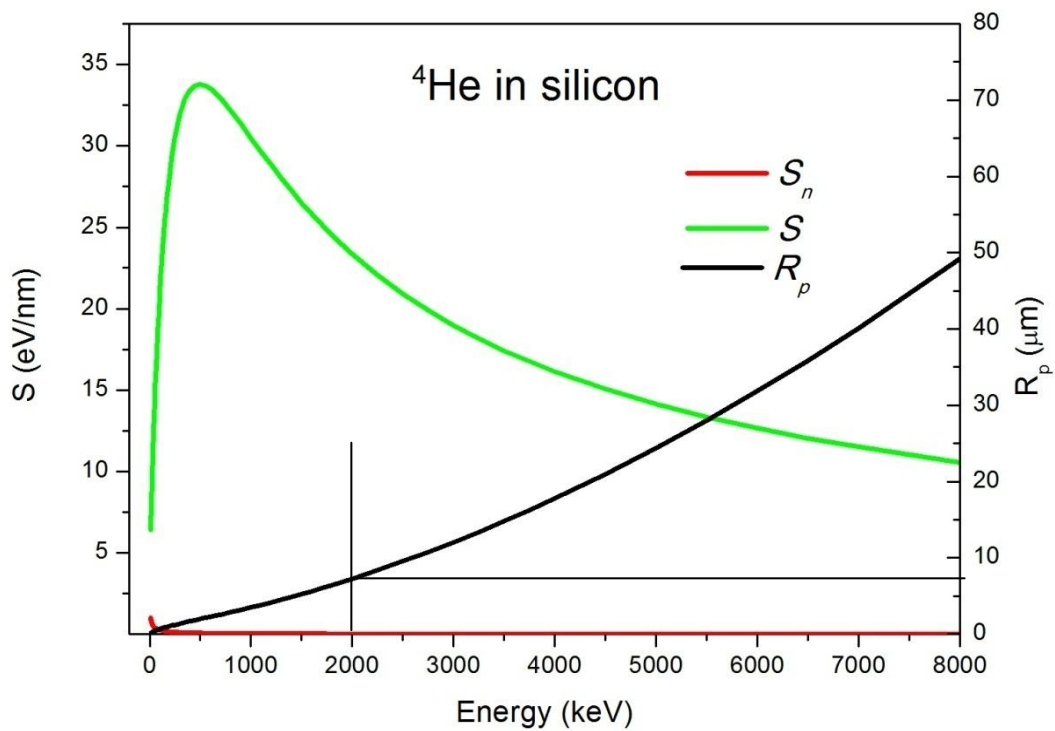

***Figure 2 :*** Stopping power and rage of an alpha particle travelling in silicon as a function of its kinetic energy

### The consequences of electronic energy loss

There may be some debate as to what proportion of energy is lost directly to collective electron modes (plasmons) and what proportion is lost in localized particle-electron interactions (Lindhard proposed an 'equi-partition rule'), but the end result is that practically all of the energy ends up in the plasma oscillations since electrons which gain energy in direct collisions rapidly also shed their energy to the collective modes. For typical electron densities, plasma frequencies are of the order of THz – so that the period is of the order of $10^{-12}$s – of the same order as the time taken for a 2 MeV

$^4$He to lose all of its energy in silicon. Thus the analyzing particle is practically already at rest before the milieu has time to respond. The plasma subsequently oscillates and interacts with the atomic nuclei, eventually setting them in motion via the plasmon-phonon interaction − essentially heating the solid. This relaxation occurs over a substantially longer time period than that of the trajectory of the particle.

There is significant inner shell ionization (giving rise to characteristic x-ray emission and Auger electron emission, for example). In conductors there is a pool of delocalized conduction electrons that are available to fall into any holes so that there is no long-lived ionization within the material. In the case of semiconductors, the density of conduction electrons is smaller than that in metals, and they need to cross the bandgap before they can neutralize any local ionization, whilst in the case of insulators there are no delocalized conduction electrons. Thus the response of metals, semiconductors and insulators to the electronic energy loss is quite different. Qualitatively, it may be observed that metals are quite radiation resistant (as for example seen in channeling studies); semiconductors are fairly resistant, and insulators can be very sensitive. In extreme cases, insulators may build up sufficient internal charge that electric fields strong enough to cause macroscopic damage are created : the sample may literally blow itself apart, or become severely cracked.

### *Organic materials*
The above discussion is based around inorganic materials, however the situation is quite different in the case of organic materials such as polymers, biomaterials and organic crystals. In these cases the structure of the materials is very sensitive to the chemical bonding. Ionisation from $S_e$ can break chemical bonds and stimulate the formation of new ones, inducing radical changes in the material's composition and chemical structure. Practically all organic materials that are subject to IBA contain significant quantities of hydrogen, and it is almost a universal observation that the breaking and reorganisation of bonds results in the release of hydrogen during IBA of these materials. The hydrogen may be released all along the ion tracks, and recombine into hydrogen molecules which diffuse and eventually find their way into the vacuum of the analysis chamber. The remaining polymer then has an increased number of double carbon bonds and in extreme cases may become graphitic. Similar mechanisms may also lead to loss of carbon, oxygen and other species from the polymer during analysis. It should also be noted that many polymers are rather heat sensitive – they are often poor thermal conductors and may either melt or undergo chemical transformations at temperatures that are low compared to the melting points of inorganic crystals. In these cases, the heating of the material by the incident beam (which is an indirect result of the energy loss in the sample) may melt or modify the polymer. The chemical changes induced in polymers during IBA under vacuum or inert gas (e.g. He in an external beam setup) may also leave them more reactive to oxygen or water vapour from the air, leading to further modification once to the material is once again exposed to air.

### *The consequences of nuclear energy loss - Frenkel pairs*
Although constituting only a small proportion of the overall energy deposition into the sample during IBA, the energy lost directly to atomic nuclei in binary collisions with them can be a major source of damage. This is because the energy lost in a single collision can be great enough to displace the atom permanently from its equilibrium position in the structure, which may be an irreversible change. This primary defect, which consists of an interstitial atom and a vacancy, is called a Frenkel pair. This is in contrast to all of the energy lost to plasma excitations and then relaxed to phonons – this may slightly heat the sample but once the sample cools down, there is a good chance that everything will return to its initial state. Let us again consider a 2 MeV ($E_\alpha$) alpha particle traversing silicon. Applying non-relativistic classical mechanics (which is adequate here) to the collision between the alpha particle of mass $M$ and a target entity of mass $m$, we may write the maximum energy transferable $E_{max}$ in a single collision as :

$$E_{max} = \frac{4Mm}{(M+m)^2} E_a.$$

For collision with an electron (m=M/7344) $E_{max}$ is 1.1 keV, whereas for collision with a Si nucleus (m=7M), $E_{max}$ is 875 keV. These events are very rare, requiring a head-on collision and the average energy losses are an order of magnitude smaller but even a 1 keV electron cannot transfer to a Si atom the minimum energy $E_{disp}$ (a few tens of eV) necessary to displace it permanently from its lattice position. The alpha particle is largely able to impart such an energy.

The study of the formation of defects due to $S_n$ is a complicated field because the real displacement energy – that energy necessary to be given to a target atom for it to form a stable Frenkel pair – depends amongst other things on the detailed atomic movements necessary to allow the passage of the displaced atom from its lattice site to the nearest stable interstitial site (including relaxation of the crystal structure as the atom is displaced) and also on the direction in which the displaced atom is initially pushed. Calculation of these quantities requires detailed density functional and ab-initio calculations, so usually $E_{disp}$ is a phenomenologically adjusted value.

### Collision cascades

Furthermore, in many cases the energy imparted to a target atom is sufficient that it may, in turn displace further atoms, which may – in their turn, also have enough energy to induce further displacements. The resultant process, in which a primary knock-on atom (PKA) initiates a sequence of further knock-ons, is called a cascade. The cascade may be characterised by the total number $n_{disp}$ of displaced atoms, the density and spatial distributions of interstitials and vacancies within the cascade, the density and spatial distribution with which cascades of given $n_{disp}$ occur along the ion track, and so on. This view of the cascade, which is the one encountered in IBA, is a linear cascade, in the sense that in principle it can be modeled as a series of binary collisions. The well-known SRIM code can be used to perform Monte Carlo calculations of such linear cascades. In figures 3 and 4, are shown SRIM (2008.04) Monte Carlo calculations of the trajectories and cascade formation for 2 MeV $^4$He particles in Si, for 2 MeV Xe in Si, and for 120 MeV Xe in Si. In the first case, after some 1000 ions, there has not yet been one significant cascade formation. In the second, after just 25 ions, it is clear that there is formation of numerous cascades for each incident ion.

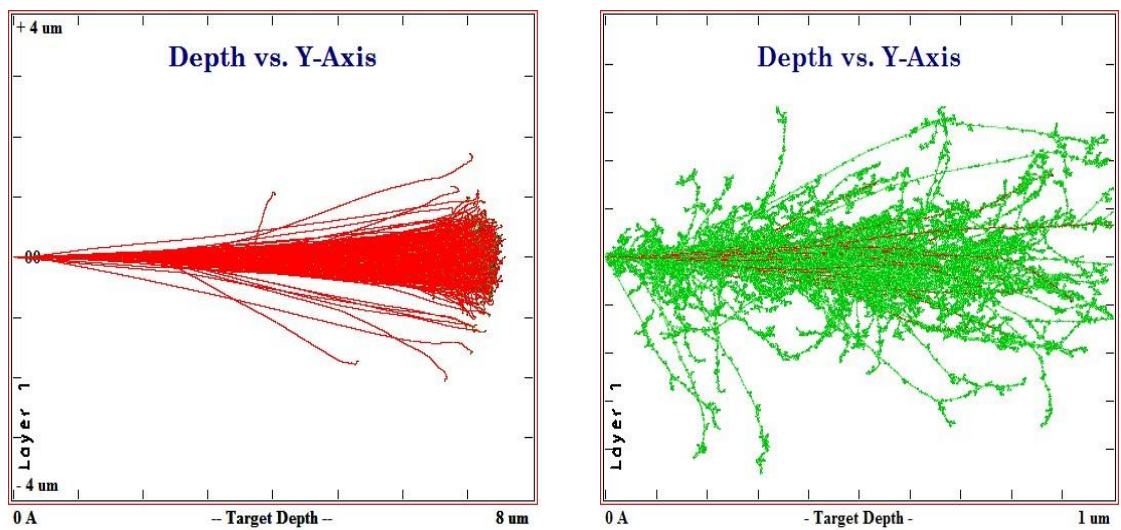

**Figure3 :** SRIM (2008.04) Monte Carlo calculations of the trajectories and cascade formation for 2 MeV $^4$He particles in Si and for 2 MeV Xe in Si

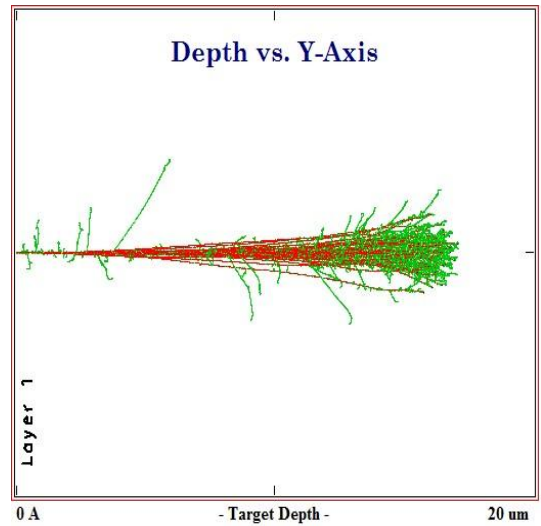

***Figure 4 :*** SRIM (2008.04) Monte Carlo calculations of the trajectories and cascade formation for 120 MeV Xe in Si.

The third example is close to the IBA situation for heavy ion high energy ERDA. There, the beam would be incident at a low angle – between 10 and 30°. The calculation is for 60 ions, and it is clear that there could be significant loss of material since at low angle beam incidence some cascades would intersect the surface.

Although Monte Carlo and more sophisticated calculations are required for detailed calculations of cascades even in the linear regime, it is possible to obtain a simple analytical expression for the average number $\bar{n}$ of displaced atoms in a cascade due to a PKA of energy $E_0$, using simple reasoning. Firstly, the maximum number of displacements cannot be greater than $E_0/E_{disp}$, in which every collision leads to a displacement by transmitting just $E_{disp}$. The minimum number is 1 – in this case no collision transfers more than $E_{disp}$ to any atom (so only the PKA counts as a displaced atom for the cascade). If we assume that all numbers between these two limits are equally probable, then $\bar{n}$ is just ½(1+$E_0/E_{disp}$) or, supposing that $E_0 \gg E_{disp}$ as is commonly the case then

$$\bar{n} = \frac{E_0}{2E_{disp}}.$$

This result was first obtained by (Kinchin and Pease 1955) and is close to results obtained by much more sophisticated reasoning. It can be used in SRIM to increase the calculation speed when calculating numbers of vacancies and displacements, at the expense of obtaining reliable cascade statistics.

### *What happens after the cascade?*

Calculations such as those performed by SRIM give the initial production of defects due to the interaction of the ion with the solid. However if we observe the solid at some time *t* much greater than the formation time (typically 1 to 100 ps) of the cascades we will not observe the expected numbers of interstitials and vacancies. This is because both during the cascade and afterwards, the defects are mobile and so a substantial number of interstitials encounter vacancies thus annihilating a Frenkel pair, or migrate and interact to form stable extended defects such as vacancy clusters, dislocation loops and so on. In polyatomic materials, an interstitial atom of type A may fill a vacancy of an atom of type B, C, D… , resulting in an anti-site defect. Because interstitials are much more mobile than vacancies (vacancies are created with no momentum, whereas interstitials may have non-zero momentum) the vacancy and interstitial populations do not share the same spatial distribution within a cascade. The interstitials tend to be more concentrated at the periphery of a cascade, whereas the vacancies tend to be more concentrated in the core of the cascade.

***Beyond the linear cascade model***

Although seldom if ever encountered in IBA, in the interests of completeness we mention here a further damage mechanism. At very high electronic energy loss rates (for example 1 GeV Pb in zirconia, where $S_e$ is around 30 keV/nm) the extremely high energy density deposited into the electronic system and then transmitted to the atomic lattice results in extreme temperatures, far from equilibrium, and phenomena such as transient melting or plasma formation along the ion path, strongly enhanced ejection of matter from the surface, formation of bumps or hollows on the surface, and generation of thermal shock waves. For some of these processes the thermal spike model (Toulemonde et al. 1992) provides quantitative or qualitative insights.

***Sputtering***

When a recoiled atom reaches the surface of the sample with a positive momentum component along the surface normal and associated kinetic energy greater than the surface binding energy $U_s$, it can leave the surface of the sample. This is the process of sputtering. The physics of sputtering has been studied extensively by Sigmund using an analytical physics approach. A most useful summary and bibliography is given in (Sigmund 2012). Calculation of sputtering yields can also be made via Monte Carlo calculations. The SRIM binary collision MC code (Ziegler et al. 1985) (http://www.srim.org) for example has provision for calculating sputtering yields, and it is both useful and interesting to explore the tutorial (http://www.srim.org/SRIM/Tutorials/Tutorials.htm) that is found on the SRIM website. One key element is the choice of the surface binding energy. A useful starting point is the energy of sublimation, and SRIM proposes this as a first choice for elemental targets, although extension to polyatomic targets (via some kind of linear combination rule) is hazardous: good experimental values are preferable.

An ultimate limit to nondestructive analysis is given by the surface sputtering induced by the incident ion beam. For a rough estimate of the minimum layer thickness that can still be detected by an IBA technique, (Feldman and Mayer 1986) have used the following consideration. For a beam spot area A the number of atoms per unit area $\rho_{sp}$ sputtered by a total number of projectiles Q is $\rho_{sp}$ = Y·Q/A, with Y being the sputter yield. The number of particles that are detected at the same time is $N_{det}$ = σ·Ω·Q·ρ, where σ is the scattering cross-section of the used analysis technique, Ω is the detector solid angle and ρ is the total area density of the layer under investigation. To first order, the sputtered layer thickness is thus $\rho_{sp}$ = (Y·$N_{det}$)/(A·σ·Ω·ρ). Requiring that the layer may not be completely removed during the measurement ($\rho_{sp}$ < ρ) yields an approximate lower limit for the layer thickness that can still be detected:

$$\rho_{min} > (Y \cdot N_{det} / A \cdot \sigma \cdot \Omega)^{\frac{1}{2}} . \qquad (1)$$

Of course, $N_{det}$ has to be defined by the requirements of the specific analytical problem. As long as electronic sputtering can be neglected simple collisional sputtering can be estimated approximately (Wang and Nastasi 2010) by

$$Y = 0.1 \cdot \frac{S_n}{U_S \cos \varphi} , \qquad (2)$$

where $S_n$ is the nuclear energy loss of the projectile in eV/$10^{15}$ at/cm$^2$, $U_S$ is the surface binding energy in eV and $\varphi$ is the incidence angle of the ion beam to the surface normal. Better estimates can be obtained from SRIM (www.srim.org) or from various semi-empirical estimates from the literature. A useful table of sputtering yields relevant for IBA is given in (Cookson 1988). We may calculate some orders of magnitude, for a standard RBS analysis. Consider the effect of sputter induced damage during 2 MeV He RBS of a 5x10$^{14}$ atoms/cm$^2$ Au layer using a 1 mm$^2$ beam spot and a detector with 1 msr solid angle. Both formula (2) and SRIM predict a sputter rate $Y$ of about 4x10$^{-3}$ for gold bulk material (which of course is not necessarily the same as $Y$ for a monolayer). From equation 1 we calculate that about 200 counts will be accumulated from a Au layer of 5·10$^{14}$ atoms/cm$^2$ before the whole layer is sputtered away. For a typical HIBS set-up using a 10 MeV Si beam with the same spot area and the same detector solid angle an estimated 150 counts can be acquired before removal of the layer. In other words, after the acquisition of a backscattering spectrum with 200 counts, the film is essentially destroyed. This is a severe limitation for the detection limits of surface contaminants under these analysis conditions (Pedersen et al. 1996; Banks et al. 1998). The type and energy of the beam used for elastic scattering does not influence this number considerably, since to first order both the sputtering of the surface and the elastic scattering are Coulomb processes so the same scaling of cross-sections applies. The effect of sputtering during IBA can only be mitigated by increasing the detector solid angle and the beam spot area or by maximizing the scattering cross-section by selection of optimum cross sections (e.g. forward scattering angles). For very thin films it is recommended to estimate the amount of surface erosion by the above formulae. (Döbeli and Müller 2011) have shown that for metal films with a thickness larger than about a nm, SRIM provides sufficiently accurate estimates of sputtering rates for this purpose.

It is to be noted that this simplistic calculation ignores elemental loss due to electronic energy deposition processes – so-called 'electronic sputtering'. This is acceptable for most IBA situations, however when heavy ions of high energy are used, electronic sputtering may cause significant material loss, as has been documented for example for heavy ion ERDA (Walker et al. 1998) in insulators.

### Beam heating effects

Almost all of the energy deposited in the sample by the analyzing beam eventually ends up as heat in the sample. The range of powers per unit area delivered to a thick sample during IBA may range from a few hundred mW/cm$^2$ for a total incident power of a few hundred mW (a few hundred nA of 1 MeV particles in a 1 cm$^2$ beamspot) to several hundred kW/cm$^2$ (e.g. 1nA of 2.5 MeV particles in a 1 um beamspot for a total power of 2.5 mW). Obviously, the temperature rise associated with the analysis may cause modification of the sample – elastic or plastic deformation; enhanced diffusion of constituents; modification of existing or accumulating crystalline damage; sublimation; chemical reactions within the sample and so on. Given the vast array of materials investigated with IBA, with a huge range of physical, chemical and thermal properties, it cannot be expected that the ion beam analyst be able to predict the effects of beam heating for each case, however it is necessary to be able to estimate at least an order of magnitude of the temperature increase expected in the sample, due to the incident ion beam. Even this is rather difficult, because the thermal properties of the samples being analysed are rarely well known, and even when they are the sample temperature rise may also sensitively depend on the thermal coupling of the sample to its surroundings: a thin self-supporting thermally insulating sample will reach quite different temperatures to those attained by the same sample thermally well connected to a large heat sink (e.g. a solid metal sample holder). It is thus useful to first examine what has been observed experimentally, in terms of sample temperature rise during typical IBA.

Reliable experimental measurement of sample temperatures during ion beam irradiation are extremely difficult, however because of the very high beam current densities that result from nuclear microprobe measurements, it was mainly from this community that the first estimates of sample temperatures came. (Ingarfield et al. 1981) considered the temperature increase at the surface of a silicon wafer due to a few tens of nA of 2 MeV $^4$He in a 10μm diameter spot, by equating the input power density with that of a CW laser for which temperature curves had been calculated (Williams et al. 1978). The steady state temperature was found to be less than 200°C. In one of the first non-microprobe studies, (Hansen et al. 1980; Hill et al. 1981) found, by irradiating wax particles with calibrated melting points deposited on thin Mylar or Kapton backings, that 150nA of 2 MeV protons in a 43 mm$^2$ spot irradiating particulate samples on a 3.2 μm mylar backing resulted in a maximum sample temperature of 225°C.

Given the difficulties in measuring sample temperatures, attempts have been made to calculate sample temperatures. Whilst the input power distribution from the ion beam may be well characterised, and in principle the heat equation solved (with numerical tools if necessary) for practically any sample geometry, there is a major problem in being sure to have valid thermal conductivities and heat capacities, and in adequately catering for realistic heat transfer coefficients between various elements. In vacuum, small mechanical imperfections between nominally thermally coupled elements (e.g. a silicon wafer clamped to a metal support) assume a greater importance than in air, where the air itself may provide improved thermal transfer between the imperfectly coupled elements. The effort required to make accurate thermal calculations in real IBA situations is probably too great to be justified, when a pragmatic approach (put in a trial sample and see what happens …) often provides an adequate answer to the question 'will my sample be damaged by thermal effects?'. Nevertheless, calculations for some idealized situations help build intuition, and give a sense for the orders of magnitude involved.

### Conductive cooling - Thin samples

The main calculations have been for irradiation of thin foils with their edges held at fixed ambient temperature, in vacuum, under the assumption that the temperature remains sufficiently low that heat loss by radiation is far smaller than heat loss by conduction to the support structure. In the course of studies of ancient papers by PIXE, (Cahill et al. 1986) considered such a thin foil system, including the case with one side cooled by a flowing fluid (such as air) and provide useful formulae – if one knows the thermal properties of the sample being analysed. (Cookson 1988) considered the same system, and calculated from simple analytical solutions to the heat equation that the maximum temperature induced in a 'biological sample' of thickness a few microns by a 100pA beam of 3MeV protons in a 1μm beamspot, is only about 12°C above ambient.

We show here the main result obtained by Cookson, for a homogeneous circular beam of radius $R_1$ incident on a thin target of diameter $R_2$ held in a radial holder, acting as an infinite heat sink held at temperature $T_2$

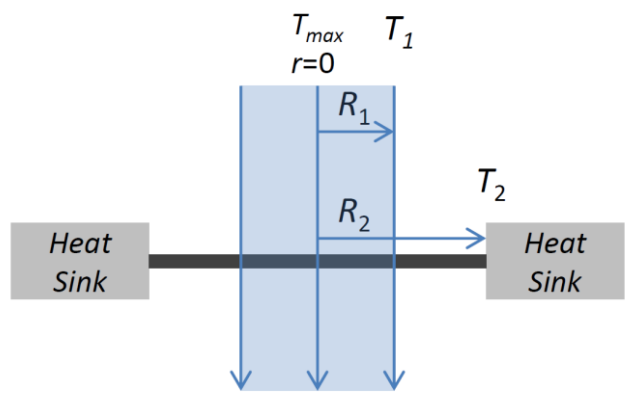

***Figure 5 :*** re-drawn from (Cookson 1988)

If the power deposited per unit volume in the region 0<$r$<$R_1$ is $q$, and the thermal conductivity of the material is κ, then $T_{max}$, $T_1$ and $T_2$ are related by :

$$T_1 - T_2 = \frac{qR_1^2 \ln(R_2/R_1)}{2\kappa}$$

$$T_{max} = T_1 + \frac{qR_1^2}{4\kappa}$$

$$= T_2 + \frac{qR_1^2 \ln\left(\frac{R_2}{R_1}\right)}{2\kappa} + \frac{qR_1^2}{4\kappa}$$

$$= T_2 + \frac{qR_1^2}{4\kappa}\left[1 + 2\ln\left(\frac{R_2}{R_1}\right)\right]$$

with the radial dependencies given, in the region $0 < r < R_1$ by :

$$T(r) = T_1 + \frac{qR_1^2}{4\kappa}\left[1 - \left(\frac{r}{R_1}\right)^2\right] \qquad (3)$$

and in the region $R_1 < r < R_2$ by :

$$T(r) = T_1 - \frac{qR_1^2}{4\kappa}\left[2\ln(r/R_1)\right] \qquad (4)$$

These relationships are plotted here :

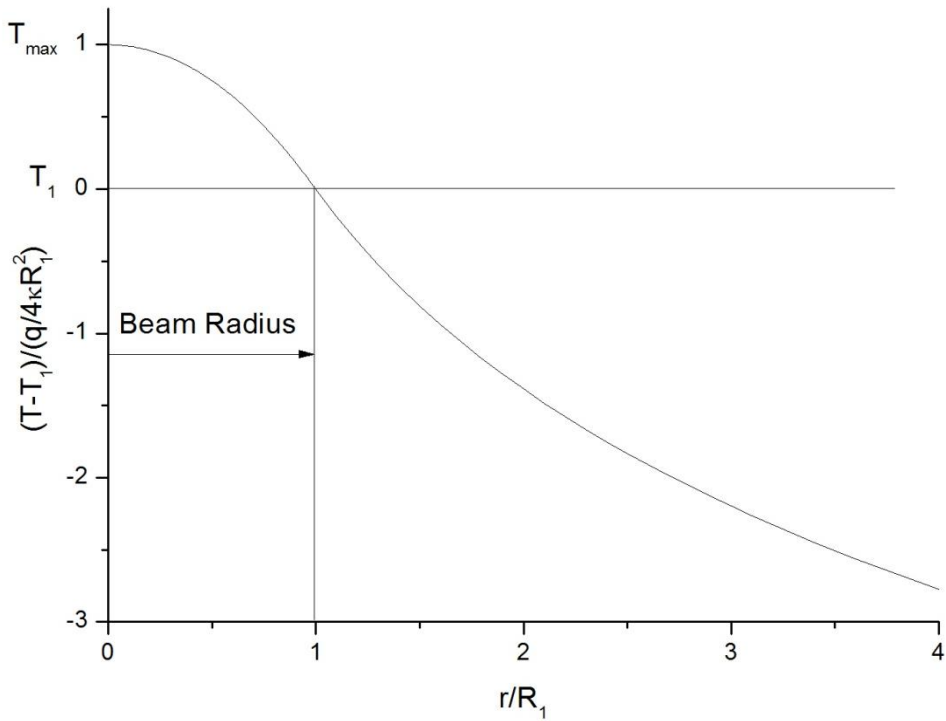

***Figure 6 :*** Equations 3 and 4 in reduced units, after (Cookson 1988)

We note that if the stopping power is constant along the path of the beam in the thin sample, then the temperatures are independent of the sample thickness since $q$ is invariant. Intuitively, this is reasonable: a thicker sample absorbs more energy from the beam, but is also better able to conduct it to the heatsink. For intermediate thickness samples – a significant fraction of the beam range for example, where the energy density delivered to the target varies significantly along the beam path, an average value of $q$ may be used. For samples thicker than the projected range $R_p$ of the ions, the energy deposition no longer increases as the sample thickness increases, whereas the conduction of heat from the beam spot to the sample surrounds does increase: the maximum temperature calculated for a thin self-supporting sample is therefore an upper limit to the maximum temperature that will occur in a thick sample – all other things being equal.

Some thermal conductivities of representative materials are given in Table 1, together with $T_{max}$ calculated for 3 MeV protons and 1 MeV $^4$He particles for a macrobeam situation (20nA in a 2mm beamspot ($R_1$ = 1 mm), in a 1cm ($R_2$=5mm) target holder and a microbeam situation (1nA in 1um beamspot ($R_1$ = 0.5 um), in a holder of diameter 100 um ($R_2$ = 50 um).

| Material $\kappa$ (Wm$^{-1}$k$^{-1}$) | $S$ (keV/μm) $R_p$(μm) | | Thin Target conduction only | | | | Thick Target conduction only | | | |
| | | | macro $T_{max} - T_2(K)$ i = 20nA, $R_1$=1mm | | micro $T_{max} - T_2(K)$ i = 1nA, $R_1$=0.5μm | | macro $T_{max} - T_0$ i = 20nA, $R_1$=1mm | | micro $T_{max} - T_0$ i = 1nA, $R_1$=0.5μm | |
| | $^1$H 3 MeV | $^4$He 1 MeV | $^1$H 3 MeV | $^4$He 1 MeV | $^1$H 3 MeV | $^4$He 1 MeV | $^1$H 3 MeV | $^4$He 1 MeV | $^1$H 3 MeV | $^4$He 1 MeV |
| Aluminium **250** | 22.5 132 | 337 3.2 | 2.4e6 | 3.6e7 | 0.07 | 1.1 | 0.07 | 0.03 | 0.03 | 0.5 |
| Silicon **148** | 19.7 92.1 | 305 3.5 | 3.6e6 | 5.5e7 | 0.11 | 1.7 | 0.11 | 0.04 | 0.06 | 0.6 |
| Gold **310** | 71.8 26.8 | 732 1.5 | 6.2e6 | 6.3e7 | 0.19 | 1.9 | 0.06 | 0.02 | 1.2 | 0.7 |
| Silica **1** | 42.0 15.7 | 328 3.6 | 1.1e9 | 8.8e9 | 34.2 | 267 | 16 | 6.4 | 6.1 | 89 |
| Mylar **0.15** | 15.5 115 | 266 4.3 | 2.8e9 | 4.8e11 | 84 | 14420 | 109 | 42 | 55 | 492 |
| Dry Biological Tissue **0.086** | 12.4 141 | 227 5.1 | 3.9e9 | 7.1e10 | 117 | 2146 | 185 | 74 | 77 | 725 |

**Table 1:** Calculated temperature rises if only conduction is invoked, for selection of experimental conditions.

The $qR_1^2$ dependence for $T_{max}$ for thin targets leads to unrealistically high values of $T_{max}$ for macrobeams as may be seen in the table and in practice heat must be lost by other mechanisms than conduction. The most important is radiative cooling, which would be expected to dissipate most of the incoming energy in this case.

### Conductive cooling-Thick samples

For a uniform beam profile delivering power $P$ is to an infinitesimally thin circular slab of radius $R_1$ in a semi-infinite solid held at $T_0$ far from the impact point, we may write for the steady state (Carslaw and Jaeger 1959; Lax 1977)

$$T_{max} = T_0 + P/(\pi\kappa R_1).$$

This equation is valid when the beam radius $R_1$ is much greater than the beam projected range $R_p$ in the solid. As $R_1$ approaches $R_p$, the maximum temperature reached is smaller than $T_{max}$ by some factor $N<1$. (Lax 1977) has also solved the heat equation for the case when the power is delivered into a depth that is non-zero – for the case of a Guassian beam power profile and an exponentially decreasing power dissipation (appropriate for lasers, but also applicable to have an order of magnitude estimate for ion beams) and gives an expression for $N$. This is shown in Figure 7 (Fig 1 of (Lax 1977)), where $W$ is $R_1/R_p$. We may note that for beam radii greater than about 10 $R_p$ $N$ is greater than 0.9, whereas for for $R_1 \cong R_p$ $N$ is around 0.5. Values on $N$ from Figure 7 are used for the calculation of the temperature rise in thick targets, due to conduction only, in the table above. For the typical microbeam situation of a beam of radius $R_1$ of 0.5μm and an $R_p$ of around 100μm (W=0.005) $N$ is clearly very small – of the order of a percent, but $T_{max}$ may be very high (e.g. 10000 °C for Mylar, and 23000°C for biological tissue), so temperature rises of some hundreds of degrees can be envisaged in thermal insulators.

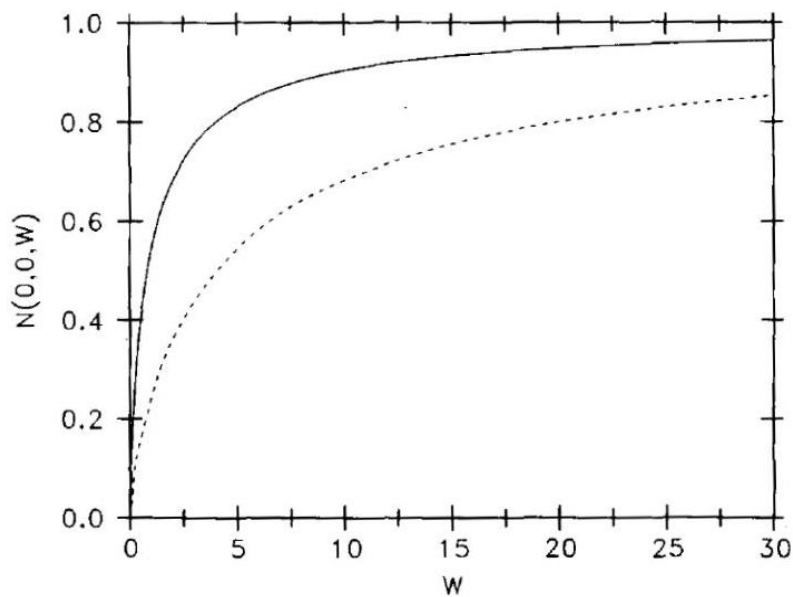

**Figure 7 :** from (Lax 1977). W on the figure corresponds to $R_1/R_p$. For the dotted line, the x axis is multiplied by 5 (so the '5' corresponds to W=1) to allow reading of N for smaller values of W.

For the typical case of a 2mm 20nA 1MeV alpha beam incident on silicon we find a temperature increase of only 0.06°C.

### Radiative cooling
Again building on the work of Cookson, we note that the heat $H$ radiated by a black body of temperature $T_{max}$, to its surroundings at $T_2$, is :

$$H = 5.7 \times 10^{-8} \varepsilon (T_{max}^4 - T_2^4) \text{ Wm}^{-2}$$

The emissivity ε varies from a maximum of 1, to about 0.6 for common materials, and to as low as 0.05 for highly polished metals. Making a crude assumption that the sample is at $T_{max}$ throughout the beam heated zone, and at $T_2$ elswhere; that the beam is (just) completely stopped in the sample; and assuming radiation from front and back sample surfaces, we may equate the incoming and outgoing energy fluxes to determine the steady state temperature :

$$T_{max}^4 = T_2^4 + \frac{1}{2} 1.8 \times 10^7 \frac{\Delta Ei}{\varepsilon A}$$

where $\Delta E$ is the average energy in eV lost by the beam in the sample, $i$ the current in amps, and $A$ the area in m$^2$. The value of $T_{max}$ is rather insensitive to assumptions about the emissivity, about the actual area $A$, and the assumption of emission from both sides of the sample, due to the 4[th] power dependence of $T_{max}$ on these quantities. For example an order of magnitude increase in $A$ (i.e. supposing that conduction leads to significant temperature rise outside the beam impact region …), leads to a reduction of less than a factor 2 in the temperature rise above $T_2$ and supposing radiation of heat from only one face of the sample increases $T_{max}$ by only a factor of 1.2.

| Emissivity | $T_{max}$ °C (macrobeam) | | $T_{max}$ °C (microbeam) | |
|---|---|---|---|---|
| | [1]H 3 MeV | [4]He 1 MeV | [1]H 3 MeV | [4]He 1 MeV |
| 1 | 351 | 206 | 10600 | 8300 |
| 0.6 | 437 | 267 | 12500 | 8800 |
| 0.05 | 962 | 736 | 25800 | 18900 |

*Table 2:* Calculated temperature rises if only radiation is invoked, for a selection of experimental conditions.

This calculation is overly simplistic, since the sample temperature will indeed surely rise outside of the beam impact area, due to conduction, and the radiating area should be characterised by a temperature distribution rather than the single parameter $T_{max}$, with the lower temperature areas radiating less power. Overall we may expect that the temperatures in the table are upper limits. Furthermore, in contrast to the case of the conduction limited process, in the radiative cooling limit the temperature WILL depend on sample thickness for samples thinner than the beam range since the power dissipated in the target depends on $\Delta E$ the proportion of the beam energy lost to the target. A thinner target will absorb less power from the beam but have the same radiative properties, and will thus have a lower steady state radiation-limited temperature.

Finally, to put things into perspective, let us calculate temperatures to compare with the experimental observations given above, remaining aware that that the idealized calculations are for systems rather different from the real ones described experimentally:

| | | | $T_{max}$ (°C) conduction limit | $T_{max}$ radiation limit |
|---|---|---|---|---|
| (Hansen et al. 1980) $T_{max} <= 225$°C | 2 MeV [1]H$^+$ 150 nA 0.43 cm$^2$ beam | 3.2μm Mylar film, giving 67 keV energy loss in the film. $R_1$ = 4mm $R_2$ = 5mm $\kappa$=0.15 Wm$^{-1}$K$^{-1}$ $\varepsilon$=0.5 $T_2$=300K | $10^{11}$ °C (!!) | 55 °C |
| (Ingarfield et al. 1981) $T_{max} <200$°C | 2 MeV [4]He$^+$ 30 nA 10μm beam | Bulk silicon wafer $R_1$ = 5μm $R_2$ = 10μm $\kappa$=148 Wm$^{-1}$K$^{-1}$ $\varepsilon$=0.5 $T_2$=300K | 160°C | 10500 °C |

*Table 3:* A selection of experimentally measured temperature rises.

*Commentary*

In the case of Hansen et al, the observation is of the temperature of the wax particles, not the mylar film. In a further paper (Hill et al. 1981) they point out that that they observe the same maximum temperature (i.e. melted wax particles) on a 50μm thick Kapton film, as they do on the 3μm Mylar backing. The wax particles are rather small (some microns or tens of microns in diameter) and in an ill-defined thermal setting (emissivity, thermal coupling to backing, geometry, conductivity etc). The measurements by Hansen et al were for the analysis of fine air particulates from air filters for pollution and environmental monitoring. If it is clear that the temperature of the backing does not increase dramatically during the measurement, it is equally clear that there is little real information either experimentally or from calculations concerning the local temperatures of individual analysed particles, since the real particles of interest may have thermal properties and thermal coupling to their environment that are quite different from those of the wax particles used to estimate local temperatures achieved during irradiation. The calculations from the simplified models above are of little value in assessing the actual temperature of the analysed particles – they allow only to assert that the temperature of the backing does not increase dramatically.

In the case of the Ingarfeld et al observations, the sample is essentially a bulk sample and so we may apply the semi-infinite formula. We find a maximum temperature increase of only 160°C, close to the 200°C calculated in their paper.

*Other Material Changes Induced during IBA*

The local (on an atomic scale) structural defects created within solids under IBA conditions are often responsible for modifications of their optical, electrical, magnetic, and chemical properties. These effects depend very sensitively on the detailed nature of the solids : the nature of the bonds holding the solid together, and the band structure, so it is difficult to make useful generalizations. Nevertheless, we may observe that properties of metals are barely changed by IBA, and may often be easily recovered by thermal annealing.

The band structure of semiconductors is very sensitive to the presence of even very low densities of defects – as may be seen by the effects of doping at levels well below 1ppm. It is best to assume that any IBA will modify electrical and optical properties of semiconductors.

As discussed above, insulators are very sensitive to buildup of internal charge, and also are very sensitive to very low levels of defects – for example very low concentration of colour centres may significantly change optical properties such as absorption spectra of insulators.

The often rather weak covalent bonds in polymers are very easily broken or re-organised by an IBA beam leading to significant chemical modification that can go as far as generation of volatile molecules ($H_2$, $H_2O$, CO etc) that may diffuse from their point of production to the free surface of the polymer and be lost by evaporation.

*Carbon deposition*

A special case is carbon deposition under the beam. Except in the rare cases where IBA is performed under well-controlled ultra-high vacuum conditions (for example a residual pressure of $10^{-10}$ mbar composed largely of $H_2$ and CO) the residual gas in the target chamber ($10^{-6}$ to $10^{-8}$ mbar) often contains significant partial pressures of organic carbon-based molecules – mostly oils from pumps or organic molecules transferred from hands when handling target chamber components without using gloves. These molecules may be cracked and deposited on the sample being analysed (Blondiaux et al. 1984).

## 1.2 Brief Overview of IBA methods from a damage production point of view

Figure 8 shows the cross-sections of representative ion-matter interactions that occur during ion beam analysis, for the case of protons interacting with a silicon matrix. We will consider IBA analysis methods firstly in order of decreasing cross-section

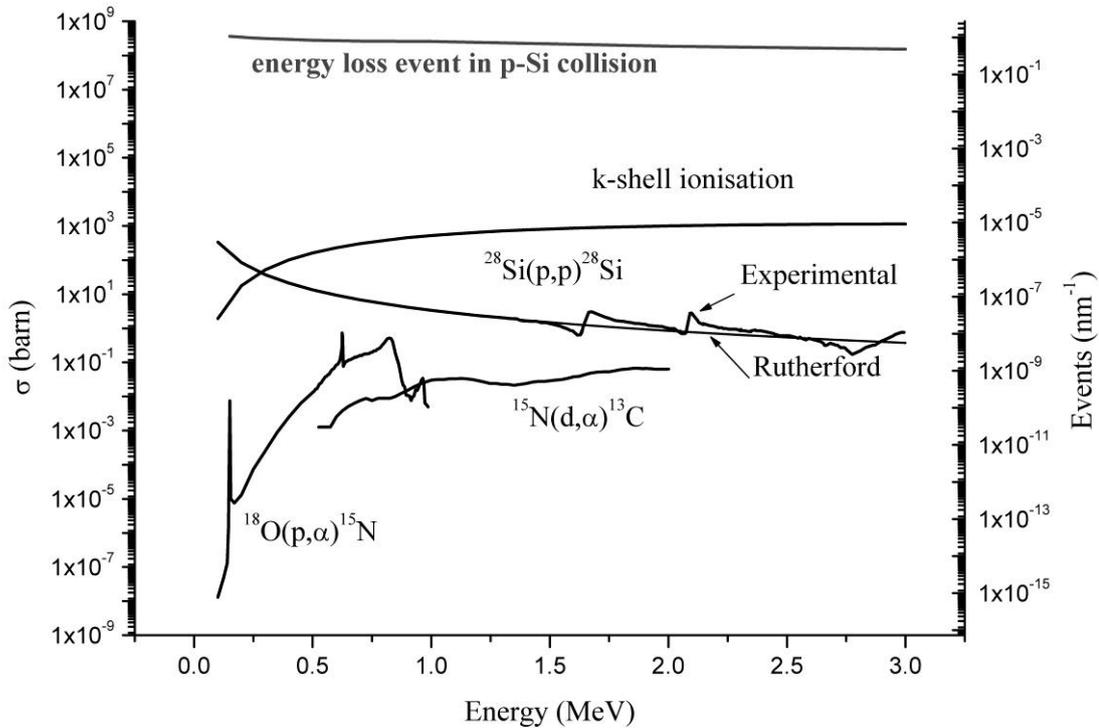

***Figure 8 :*** from (Vickridge 2004). Representative cross sections and event densities for the main ion-matter interactions occurring during IBA.

**Electronic energy Loss**

The most probable interaction by several orders of magnitude is an energy loss event in a collision between the proton and a silicon atom (for our purposes here we sum energy losses to collective electron modes and energy losses to electrons localised at the silicon atoms). This loss – the stopping power – is at the heart of all IBA concentration depth profiling methods. There is about 1 such event per nm of proton path length, and if we assume Lindhard's equipartition rule for energy losses to localised electrons and to collective modes, then this means that about 1 atom in seven along the proton's path is ionised. The electronic system rapidly relaxes and neutralises the ionised atom but clearly very special care needs to be excercised during IBA of any system for which even transient ionisation of atoms constitutes damage to the sample (for example by allowing re-organisation of chemical bonds before the neutralisation can take place).

**PIXE**

The next most probable interaction is inner shell ionisation (here we have plotted the cross section for k-shell ionisation). The resulting characteristic x-rays that are emitted when the deep hole is filled by an electron are detected in Particle-Induced X-ray Emission (PIXE) analysis. A useful introduction to the basic physics underlying PIXE can be found in (Mitchell and Barfoot 1981).The relatively high cross section for PIXE, together with the low level of interfering primary beam Brehmstrahlung radiation emitted by the decelerating protons (compared to the much lighter electrons, used in electron microprobe analysis, for example) and spectrometer resolution allowing resolution of characteristic x-rays from each element mean that PIXE has been widely used for multi-element trace element analysis. Such capability has found extensive application in analysis of materials in geology, environmental sciences, biology, and for cultural heritage materials - classes of materials that are often sensitive to damage induced by the analysing beam. Furthermore, because of the relatively high cross section, PIXE has also been favoured for use with nuclear microprobes where finely focussed beams are formed by collimation and focussing, at the expense of beam current. Although the beam current can be very small (sub nA) the local flux is nevertheless very high and here again, beam-induced damage is a major preoccupation of the analyst.

## Elastic Scattering

Moving down through the cross sections, we come to the elastic scattering of an incident particle from an atomic nucleus in the material under analysis. This process is exploited in a number of IBA techniques. When the collision is dominated entirely by the electrostatic potentials of the incident and target nuclei (we speak of Coulomb scattering) the cross section is that deduced by Rutherford from scattering of alpha particles from gold foils. If the backscattered particles are used for analysis then the method is the well-known Rutherford Backscattering Spectrometry (RBS). For energy ranges where the nuclear forces start to perturb the interaction, the reaction may remain elastic but deviations occur from the Rutherford cross section. This is visible in the $^{28}$Si(p,p)$^{28}$Si cross section on Figure 8. In this case the term Elastic Backscattering Spectrometry (EBS) is often used, and sometimes simply Backscattering Spectrometry (BS) is used to refer to both. When the recoiling target nucleus is detected (at forward angles) the technique is referred to as Elastic Recoil Detection Analysis (ERDA) or occasionally as Forward recoil Spectroscopy (FRS). In all of these cases the analysis recoils that particular atom responsible for the analytical signal from its equilibrium site in the matrix and this most certainly constitutes damage, however the success of these methods stems from the fact that the fluence of analysing beam needed for a successful analysis can usually (but not always) be kept sufficiently low that the probability of an incident particle producing an analytical signal from a previously damaged region is negligible. Although standard semiconductor charged particles are extensively used in RBS and ERDA with light ions, it is worth noting that ERDA using heavy ions often makes use of time-of-flight and/or Delta-E/E detectors, which usually implies smaller detection solid angles. In addition, higher depth resolution may be obtained by using higher resolution detectors such as magnetic ('High Resolution RBS') or electrostatic (Medium Energy Ion Scattering) detectors, which may also present smaller solid angles than for standard RBS. Elastic scattering is widely applied to metals and semiconductors, which are generally less sensitive to beam damage than the samples typically investigated by PIXE. For RBS the typical beam dose required to obtain 1000 counts (2 MeV alpha particles for 1x10$^{15}$ at/cm$^2$ Bi) is 10 μC.). Elastic scattering is often used in conjunction with PIXE, where the BS spectrum provides additional information about the sample matrix which can be used to improve the quantitative interpretation of the PIXE spectra. In such cases, the required beam dose is that for the PIXE analysis, which may result in poor statistics for the BS spectra. Even so, the extra information, even with large uncertainties, may still be very useful (Jeynes et al. 2012). One special point to note is that Type I damage is often encountered in channelling studies, where the interactions of the incident beam with the (single crystal) sample are particularly sensitive to damage. This is treated in section 2.2.

## Nuclear Reaction Analysis

The lowest cross-sections exploited in IBA are those of nuclear reactions between the incident beam and the atomic nuclei of the material being analysed. There are many nuclear reaction mechanisms, and the nuclear potentials that mediate them are not always well-known. In addition the wave nature of the particles manifests itself here, so effects such as interference and tunnelling play a role. As a result, nuclear reaction cross sections can have quite complex variations with collision energy (including resonant structures such as those manifested in the narrow peaks of the $^{18}$O(p,α)$^{15}$N cross section of Figure 1.2.1) and with the angle at which a particular nuclear reaction product is detected.

Because of the low cross sections, NRA is usually reserved for quite resistant materials. It has been extensively applied in studies of thin oxide growth, metal surface passivation, and the narrow resonances have been used in isotopic tracing studies of thin film growth. $^{12}C(d,p)^{13}C$ has a particularly high cross section, and can be used to determine the thickness of polymer thin films however damage mitigation strategies are almost always needed in this case (see section 3). It is worth noting that the cross sections between light elements of hydrogen and helium are sufficiently high that they are finding increasing application in the study of hydrogen (and deuterium and tritium) and helium interactions with cladding materials in fission nuclear reactor components and tokomak first wall cladding materials.

# 2. Minimization and mitigation of IBA Damage in different classes of materials

### Introduction

In this section, we consider the incidence and types of beam damage that occur in practical Ion Beam Analysis of various classes of materials. Each class of materials is discussed using case studies and examples drawn from the literature or from experience in the SPIRIT partner laboratories. Such a discussion cannot be complete, however the aim is to leave the reader with a practical feeling for what types of damage might be expected in 'real life' IBA situations. The bibliography contains many entries that are not directly referenced in the text : a browse over the titles in the bibliography may be helpful in finding specific information about beam-induced damage in specific materials.

### 2.1 Damage induced during IBA of metals

### Overview

Since metals have a high thermal and electrical conductivity they are generally less prone to ion beam induced damage. In addition, thin film or bulk metal samples are often polycrystalline and therefore not as much attention has to be paid to lattice defects as e.g. in semiconductor or optical materials which are prepared in epitaxial or monocrystalline form much more frequently, although important changes of grain structure under ion bombardment has been observed (e.g. Atwater et al. 1988)). In the following, an overview of possible ion beam damage mechanisms in metals is given.

### Macroscopic Damage

Metal samples will not electrically charge up under the analyzing beam to any significant degree. Therefore, macroscopic damage due to discharges (electrofracturing) or Coulomb-explosion (as possible e.g. in insulating powder samples) is not to be expected.

Macroscopic heating of the sample will always take place to some degree. As elaborated in chapter 1, the primary beam heating takes place within a surface layer of a thickness on the order of the particle range (micrometers). For a bulk metal sample the heat is dissipated relatively fast through the whole piece of material. With a good thermal contact to the sample holder the whole system can serve as a thermal reservoir and will warm up only slowly, as long as the beam power is on the order of a Watt or below. Example: 100 nA of 10 MeV particles have a heating power of 1 Watt. 100 g of aluminum would heat up by 1 K maximum in 90 seconds if thermally completely isolated. Thus, such a system will warm up by less than 40 K during a one hour experiment. Under these conditions the sample temperature can be estimated by a pyrometer through a window in the vacuum chamber. By sticking a temperature label (thermochromic pigment sticker) to some part of the sample holder the maximum temperature ever reached during the experiment can be determined.

For thin layer metal targets on badly or non-conducting substrates the heat contact to the substrate and the conductivity of the substrate material become important, as outlined in chapter 1. The local heating of an isolated metal surface layer can lead to structural changes, cracking or delamination due to thermal stress or finally to melting and evaporation.

For a substrate with bad thermal conductivity the modification of the substrate can often become more important than the damage to the metal film under investigation. Figure 8 shows an example of the damage done to a rewritable CD sample during 20 MeV Si HIBS analysis of the phase change metal coating. With a beam spot of about 1 mm diameter the polycarbonate substrate melted within approximately one minute and the metal film ruptured. By defocussing the beam to an area of about 0.5 cm2 the melting was prevented. The polymer substrate was carbonized to a large degree, but the metal layer stayed intact to a degree that made analysis feasible. By subdividing the measured spectra into several time intervals it was possible to prove that no substantial change was done to the alloy composition or film thickness.

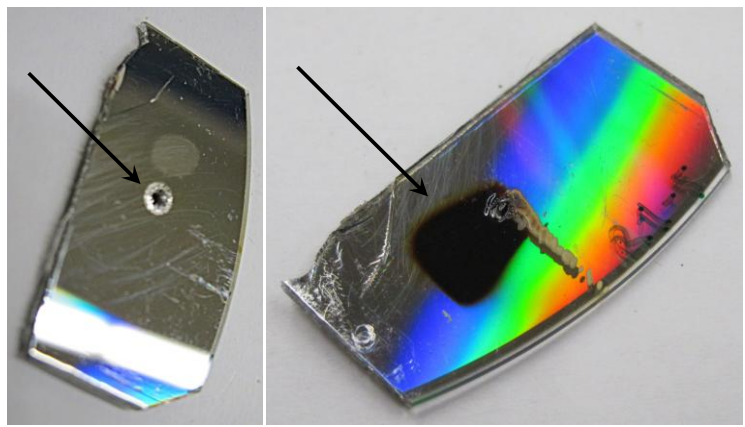

***Figure 8 :*** Damage produced during HIBS analysis of a CD-RW phase change metal coating. Left: focused 20 MeV Si beam. Right: defocused beam.

Another form of macroscopic damage can be produced by implantation of the analyzing beam particles. In metals the formation of helium bubbles at the end of range of an RBS ion beam is well known. If the helium fluence in the beam spot is excessive the surface can form blisters and the sample can virtually be destroyed. In default of an example with a metal sample, an analog case of a layer system on GaAs is shown in Figure 9. After a prolonged 2 MeV He RBS measurement with a beam spot of 1 mm diameter the sample was exposed to flash annealing to remove the suspected beam damage. This resulted in a catastrophic blast away of the sample surface down to the end of range of the He beam.

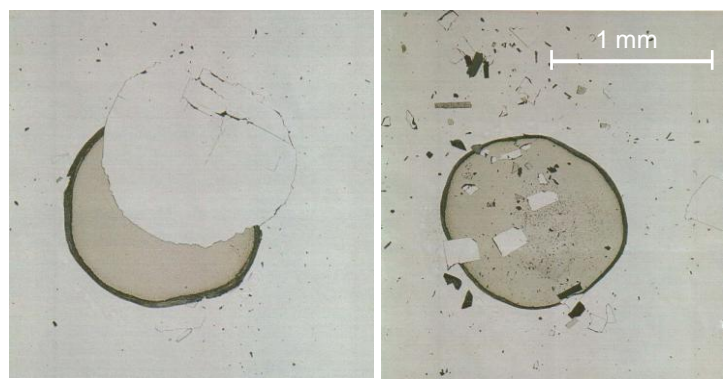

***Figure 9:*** Massive damage done to a thin film sample due to helium bubbles forming in the substrate during RBS analysis.

*Microscopic Damage*

Macroscopic thermal damage to a metal sample happens as a consequence of the thermal equilibrium state of the analyzed material or substrate and can be avoided by simple reduction of the total beam power or the beam energy flux. Time scales of macroscopic damage evolution are seconds or longer (see section 1). On the other hand, microscopic damage on the atomic scale is produced by the instantaneous physical nonequilibrium processes with time scales of ps to ns induced along the ion track by nuclear and electronic stopping forces. At low fluence, this damage is manifested in micro-structural changes with spatial extensions on the order of the track size. Ballistic displacement of atoms by nuclear collisions is inevitable during IBA in all types of materials. The primary rate of displacements can be estimated relatively well (e.g. by simple SRIM calculations). However, especially in metals, a large fraction of displacements spontaneously recombine within very short times and generally only a few percent of the primary defects survive and form more complex damage structures such as dislocation loops (Myers 1980; Wiedersich 1991a; Averback 1994; Averback and de la Rubia 1998). In metal film systems, though, damage at interfaces has to be expected caused by ballistic mixing and radiation enhanced diffusion due to the lowering of activation energies along dislocations, as these defects provide paths for rapid migration(Myers et al. 1976; Bolse 1994).

The effect of electronic energy deposition and nonequilibrium heat transients in nuclear collision cascades (thermal spike (Averback and de la Rubia 1998)) is strongly material-dependent and much more difficult to estimate (see chapter 1). Compared to other materials, metals are generally less sensitive to electronic stopping effects due to their high electron mobility (high plasmon frequencies), fast heat dissipation, and very inefficient energy transfer from the electronic system to atoms. Therefore, electronic energy loss results in remaining damage in metals only at very high stopping powers (Dollinger et al. 1996) where strong local heating can even lead to transient liquefaction and amorphization. Metallic elements and alloys that show allotropic phases are more susceptible to this kind of damage process since the material can be "locked" in a structurally or energetically close-by state (Legrand et al. 1992). In many cases, annealing of existing defects by high electronic stopping can also be observed in metals (Legrand et al. 1992), as the intensive local heating can result in a "softening" of the crystal lattice. In the present context, this has to be considered as a possibly unwanted modification of the microscopic sample structure.

At high fluence, the continuous accumulation of microscopic damage can result in changes of the macroscopic properties of metals such as density, mechanical strength, etc. (M.T. Robinson 1975).

*Impact on the IBA results*

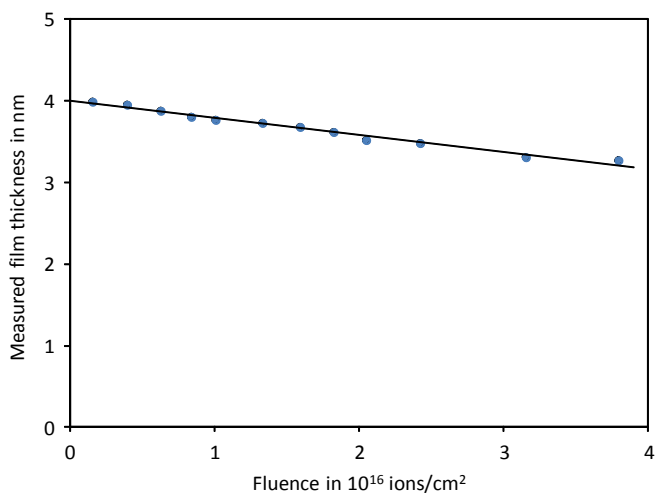

Compared to most other types of materials, ion beam analysis gives rise to only modest modification of metal samples. As a consequence, the impact of beam damage on IBA results is normally very moderate as well. Microstructural transformations will often not have visible effects in any IBA technique, except for channeling. Nevertheless, diffusion, ion beam mixing, sputtering, melting and evaporation will possibly result in a continuous or sudden change of experimental spectra. As with any other type of materials it is therefore wise to subdivide the acquired data in several time intervals to check the stability of the spectra with time. This often makes a back-extrapolation of results to zero measurement time possible (Figure 10) and, in the case of nonsystematic behaviour of experimental spectra with fluence, allows to decide if the data has to be discarded completely.

*Figure 10:* HIBS measurement of a 4 nm Cu film on aluminum with a 2 MeV Si beam. Data is divided in 12 subsets and the remaining film thickness is plotted as a function of integrated Si beam fluence. The material loss by sputtering and the pristine film thickness can be determined by a linear fit.

In Figure 11 the effect of melting on the backscattering energy spectrum of a thin metal film on silicon is shown. The Cu film was stable up to a total beam power of 1 W with a spot area of 1 mm$^2$ and started to melt when the beam power was raised to 2 W, corresponding to an energy flux of about 200 W/cm$^2$. Droplet formation led to a distinct change in peak position and shape. Using a technique with less depth resolution, the effect would probably not have been detectable.

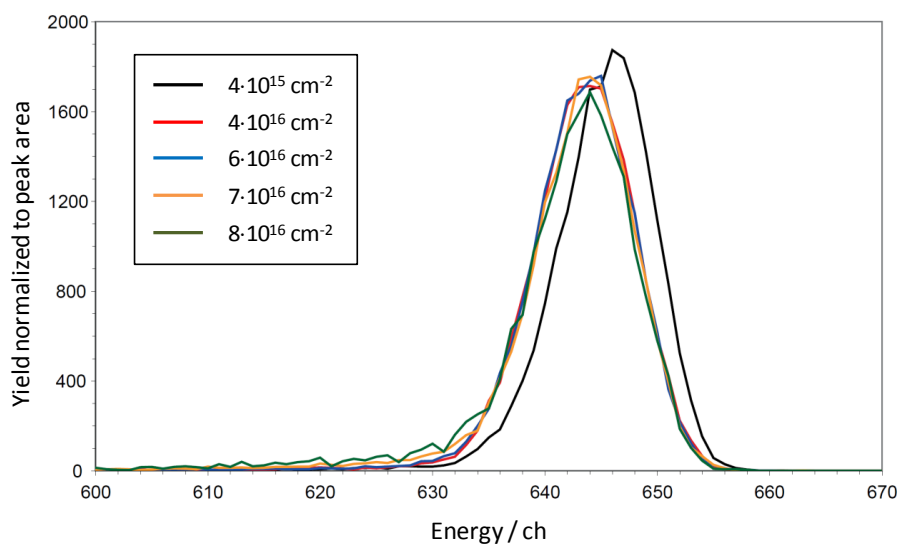

**Figure 11:** Cu signal of a 1 nm Cu film on aluminum substrate analyzed by 15 MeV Si HIBS for 5 different fluences. The peak is shifted to lower energy and a low energy tail develops with increasing fluence.

The destruction caused by overheating in the beam spot is shown in Figure 12. The metal layer has disintegrated into small droplets. Considering the big structural change of the film, the effect on the backscattering spectrum (Figure 11) is astonishingly small. No material has been removed from the surface and therefore the average areal density is unchanged. As in the previous example given in Figure 8, the broad-beam measurement of film thickness and composition yields a virtually correct result, but the sample structure is destroyed to a large extent.

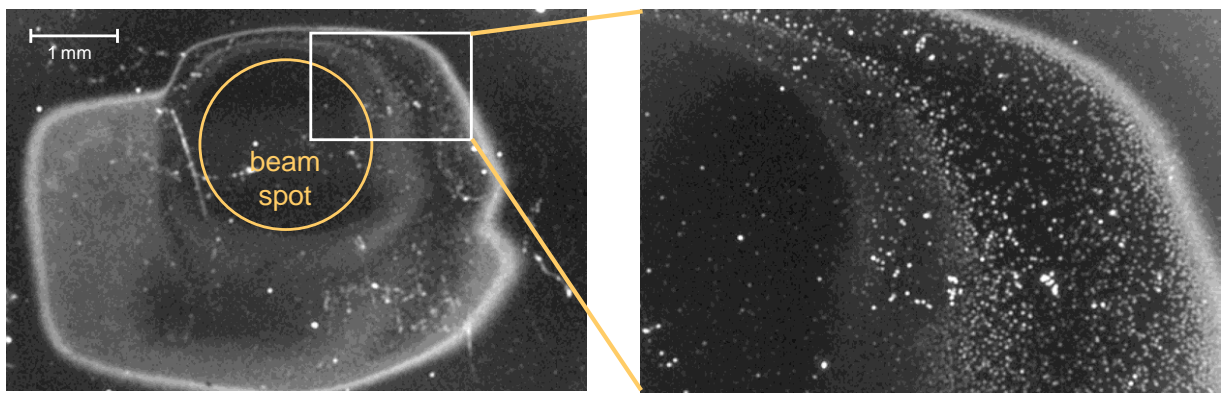

**Figure 12:** Optical micrograph of the thin Cu film on silicon after HIBS analysis described in Fig. 4. At a total beam power of 2 W in a beam spot of approximately 1 mm2 the film melted and metal droplets formed on the surface (right).

As long as only the microscopic structure and not the composition of the sample is changed, broad-beam techniques with reduced depth resolution (e.g. PIXE or non-resonant nuclear reactions) will not be dramatically affected. This can be completely different for high resolution measurements such as ERDA.

Staining of the sample surface is always a reason to worry about material deposition (e.g. by ion beam decomposition of hydrocarbons in the residual gas), removal of material (sputtering) or structural modifications that change the optical reflectivity of the material. Figure 13 shows the stain caused by a 10 MeV Ti beam on a metal film on graphite during HIBS analysis. In the beam spot less than 0.5 nm of metal was removed during analysis, nevertheless the change in surface appearance is well visible and the effect on the thickness measurement by IBA is very significant.

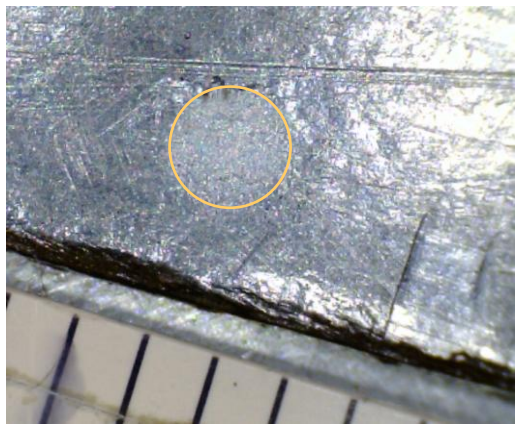

***Figure 13:*** Visible beam spot caused by sputtering Au(0.2) / Nb(0.6) / Si(3.3) nm on graphite during HIBS with 10 MeV Ti beam.

If beam damage is mediated by electrostatic charging, IBA measurements can be affected by discharges and associated base-line distortions in detector electronics. For metal films on nonconducting substrates this hazard can be avoided by contacting the metal film to the target holder.

### *IBA limits in metals*

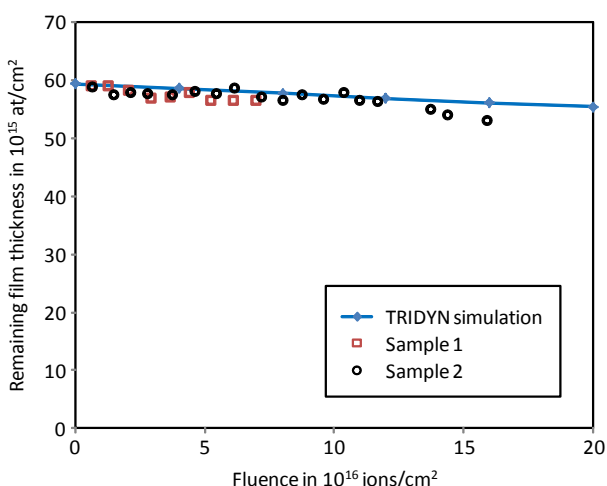

Since metals belong to those sample materials least affected by MeV ion beams, the general precautions mentioned in chapter 1 (proper selection of beam parameters, restriction of particle fluence, increase of beam spot area and detector solid angle, prevention of excessive beam heating, etc.) are normally sufficient to minimize beam damage to the sample.

***Figure 14:*** HIBS measurement of 7 nm Cu films on silicon with a 15 MeV $^{16}$O beam. Data is divided into subsets and the remaining film thickness is plotted as a function of integrated beam fluence. Comparison with a TRIDYN simulation shows satisfactory agreement with experimental data.

As mentioned before, ion beam mixing, recoil implantation and sputtering cannot be avoided and impose important applicability limits to IBA techniques. Since metals show neither chemical sputtering nor surface erosion by Coulomb explosion, the estimate of material removal by sputtering given in chapter 1 yields good results. For more precise predictions simulation programs like SRIM or TRIDYN can be applied (M. Döbeli 2011). Figure 14 shows a comparison of the measured material removal rate from a thin Cu film during HIBS analysis with the corresponding TRIDYN simulation.

This sort of simulation is generally not accurate enough for correction of experimental data but is helpful for a prediction of expected damage and optimization of beam parameters during analysis. In Table 1 some examples of material loss from thin metal films on graphite substrates during HIBS analysis are compiled as a guide.

Since microstructural damage by electronic energy loss is induced to metals for very high stopping powers only, attention to such effects has to be paid for beam particles with an electronic energy loss exceeding approximately 10 keV/nm. Again, this type of damage does not necessarily influence the IBA results but has to be considered if analysis techniques that are sensitive to the microscopic structure such as XRD or TEM are applied to the specimen later on.

| energy MeV | projectile | metal in top layer | thickness in nm | sputter yield (atoms/projectile) | | loss per 300 cts from 1 nm layer |
|---|---|---|---|---|---|---|
| | | | | measured | SRIM | |
| 15 | $^{28}$Si | Nb | 0.6 | 0.027 | 0.016 | 0.17 nm |
| 15 | $^{28}$Si | Sb | 0.7 | 0.039 | 0.031 | 0.58 nm |
| 15 | $^{28}$Si | Au | 0.2 | 0.014 | 0.028 | 0.01 nm |
| 10 | $^{48}$Ti | Nb | 0.6 | 0.15 | 0.069 | 0.11 nm |
| 10 | $^{48}$Ti | Sb | 0.7 | 0.18 | 0.18 | 0.69 nm |
| 10 | $^{48}$Ti | Au | 0.2 | 0.010 | 0.10 | 0.002 nm |

*Table 4:* Compilation of measured and calculated sputter yields from metal layers on graphite under steady state conditions. Beam incidence is perpendicular to the surface. The last column indicates the expected thickness loss during a HIBS measurement acquiring 300 counts from an initially 1 nm thick layer.

## 2.2 Damage induced during IBA of semiconductors

The energetic ions impinging onto a semiconductor during ion beam analysis can induce various types of defects, e.g. point defects, extended defects, etc. In contrast to metals, which are typically very resistant to irradiation damage by (typically high-energy light) ions, a significant fraction of the defects generated during irradiation of semiconductors do not recombine. Hence, as the fluence of impinging ions increases, an augmenting concentration of retained defects is obtained. It is important to realize that the structural defects induced by the ion beams, most often modify the functional response of the sample in turn, such as the electrical, optical or magnetic properties.

One of the simplest experiments to investigate to what extent an IBA measurement creates damage that influences the outcome of the ion beam analysis itself, is to repeat the same measurement a number of times. Such experiments have been performed to determine the damage induced by 1.57 MeV He$^+$ ions on the elemental semiconductors Ge and Si during an RBS/C experiment. Repeatedly, a channeled and a random spectrum of 30 μC (~2×10$^{16}$ at/cm$^2$) each were collected for Ge and for Si. From the channeled spectra, one can follow the crystallinity as a function of irradiation fluence. As can be seen in figure 15 (a), there is a clear impact of the He ion beam on the Ge crystal, creating enough (retained) vacancy- and self-interstitial-related defects for the minimum yield to increase with each measurement. It is also striking to note that the influence of the He beam on the Si sample (figure 15 (b)) is much smaller and almost negligible, compared to Ge. This is in agreement with the observation that Ge is much more radiation-sensitive and hence also easier to amorphize than Si. These results indicate that the IBA-induced damage is strongly related to the material characteristics, and can already influence IBA measurements during the first measurement for radiation-sensitive semiconducting materials.

Besides the defects generated within the semiconductor, another feature that needs to be taken into account during IBA measurements is the possibility to contaminate the surface with carbon atoms, when not using oil-free vacuum pumps. In these particular examples of RBS/C measurements on Ge and Si, when the samples have been irradiated for relatively long times, a C signal appears superimposed on the Ge and Si backscattering signals. Due to the energy deposition of the impinging He ions, residual hydrocarbon molecules are cracked at the Ge or Si surface, which allows carbon atoms (e.g. from the carbon-rich vapours of the vacuum system) to terminate the loose bonds at the surface.

When assessing the induced damage with ion channeling, one measures the displacement of substrate atoms from their lattice row, i.e. in the direction perpendicular to the incoming ion beam (which means, most often, parallel to the sample surface). Another approach to study the IBA-induced damage is by means of high resolution X-ray diffraction, i.e. determining the interplanar spacings (hence, mostly in the direction perpendicular to the sample surface) and deviations thereof, i.e. elastic strain. In Figure 16, several symmetric 2θ-ω scans around the surface direction are shown before and after a typical RBS/C experiment, i.e. after irradiation of 60-100 μC/mm$^2$ 1.57 MeV He$^+$ ions: (a) (004)-diffraction of pristine Ge, (b) (004)-diffraction of Co-implanted (160 keV, 2×10$^{12}$ at/cm$^2$) Ge, (c) (0002)-diffraction of InN grown on GaN on sapphire. These three examples indicate how the strain distribution in the sample can be significantly altered by a typical RBS/C experiment. Consequently, when investigating 'vulnerable' semiconductors such as germanium, it is best practice to perform the X-ray studies before subjecting the samples to ion beam analysis.

So far, we have focused on structural defects, which originate primarily from ballistic collisions between the ions and the substrate atoms. However, one should realize that any defect to the lattice, hence breaking the lattice periodicity, results in a change of the band structure of the semiconductor. Consequently, all functional properties related to this band structure, be it electronic, optical, magnetic... are likely to be modified. For example, it is well known that irradiation effects can be used to tune the transport properties of a semiconductor. The magnetic response of a dilute magnetic semiconductor can be tailored as well. In both studies, He ions with an energy of the order of MeV were used, indicating that very similar effects take place during RBS measurement.

Group III nitrides constitute a more peculiar response to the ion beam. These materials, crystallizing in a wurtzite lattice, are very radiation resistant. Indeed, even after repeatedly analysing such samples with RBS and/or channeling (similar as the cases shown in Figure 15 for Ge and Si), virtually no change in the channeling minimum yield can be observed – a very low value of the order of 1-2 % is found, indicating an excellent crystalline quality. On the other hand, even a brief RBS measurement leaves a clear coloured spot on the nitride layer, as shown in Figure 17, indicating significant optical damage. Luminescence measurement show that the irradiated part of the nitride is optically dead – no luminescence is observed at all, be it with photoluminescence, cathodoluminescence... This indicates that either very small displacement of the host atoms and/or a very low concentration of a specific defect (in both cases below the detection limit of ion channeling) are responsible for the optical defects.

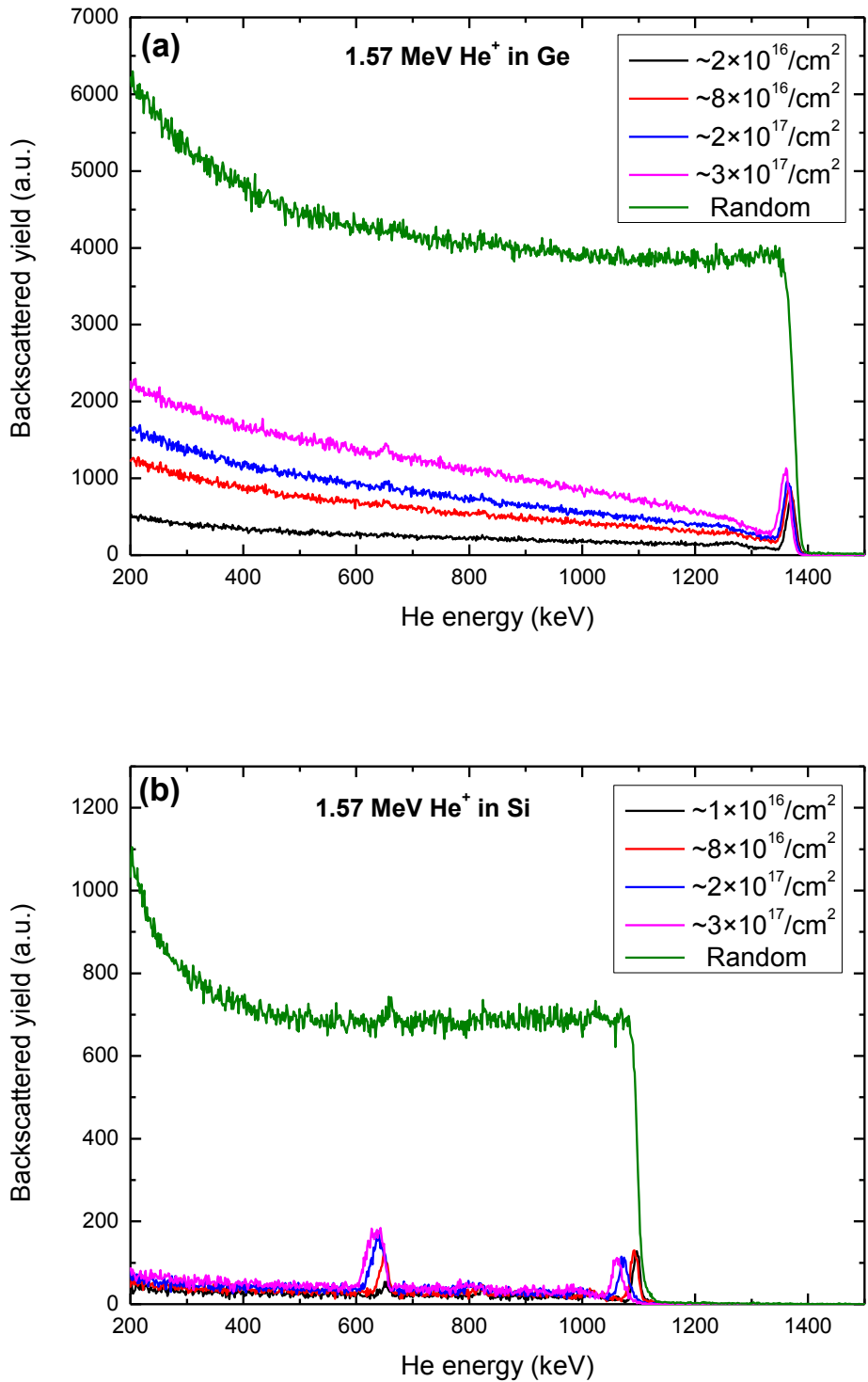

**Figure 15:** The random and channeled He backscattering signal of bulk Ge (top) and Si (down) for several 1.57 MeV He+ irradiation fluences.

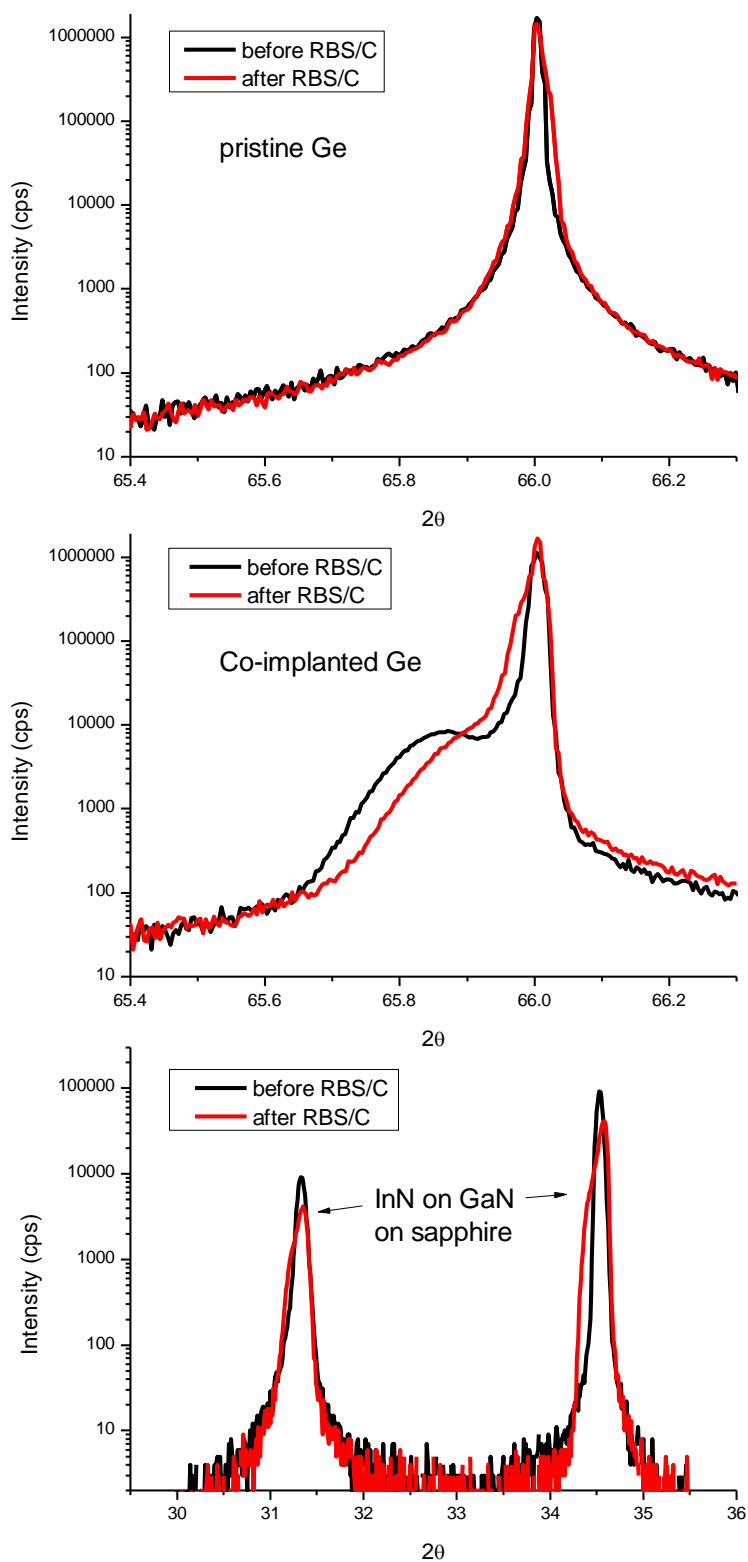

**Figure 16:** 2θ-ω scans around the (004)-direction of (a) pristine Ge, (b) Co-implanted (160 keV, 2×10$^{12}$ at/cm$^2$) Ge and (c) the (0002)-direction of InN grown on GaN on sapphire, before (black) and after (red) a typical RBS/C measurements, using 60-100 μC/mm$^2$ 1.57 MeV He$^+$ ions.

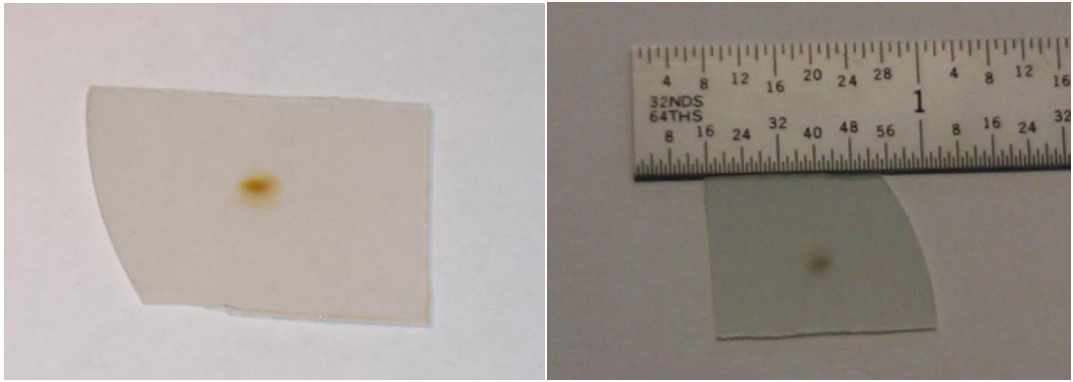

**Figure 17:** A GaN layer grown on a sapphire substrate, after RBS/C investigation with 1.57 MeV He⁺ ions. The beam spot is clearly visible by the naked eye, revealing optical damage to the nitride.

## 2.4 Damage induced during IBA of polymers, biomaterials and cultural heritage materials

### Introduction

Polymers are soft materials relatively easy to damage. Ion beams can be very damaging: even for strong materials like metals (Figure 18) the power has to go somewhere and do something! If even metals have to be treated with care then how much more polymers?

In ion beam analysis (IBA) we usually ignore the implantation aspect of the analysis. Of course the beam is implanted in the sample but the analyst doesn't care about the deep part of the implantation, where the beam has little energy and from where negligible analytical signal will come.

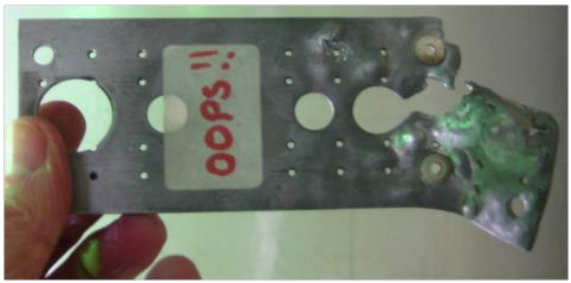

**Figure 18: Ion Beam Damage.** Energy deposition effects of a high power beam (200 keV, 0.2 mA)

Nevertheless, the nuclear scattering cross-section (highly damaging!) goes up dramatically as the beam energy falls (the Rutherford cross-section is proportional to $1/E^2$). So if your sample is thin (thin enough to not stop the beam), surprisingly at first sight, less beam damage is done, because not all the energy is deposited in the sample. This is why we can use ultra-thin (and very delicate) carbon foils as stripper foils in tandem accelerators. This is why we can use thin (8 μm) Kapton foils as vacuum windows on external beams.

Obviously, in implantation very large beam currents can be used; large, that is, from the perspective of ion beam analysts. The "oops" moment shown in Figure 18 displays a situation one imagines never met in IBA. Never? Figure 18 shows an energy density approaching 1 MW/cm³, but a nanobeam with an ultra-bright ion source might put 100 pA into a 200 nm spot: this is about 100 MW/cm³ (at 4 MeV)!

When an ion beam strikes polymers there is an immediate extra cross-linking (evolving H and other molecular fractions like OH, CO etc): this is the basis for the standard cross-linking process using electron beams (Clough 2001; Muratoglu et al. 2001). In the semiconductor industry, photoresist is a standard masking material: in implantation processing, when the beam is turned on the vacuum may collapse due to photoresist outgassing. This is a major problem for implantation technology with several different important aspects (Okuyama et al. 1978; Orvek and Huffman 1985; Wu and Kolondra 1991). The fact that your computers work at all witnesses to the fact that the problems are solved.  But the literature is large – we have merely indicated some useful entry points. We will explore various aspects of damage relevant to IBA below.

### Electronic sputtering

Beams used for MeV ion beam analysis (IBA) are almost invariably relatively high energy, and the sputtering due to the atomic displacement cascade is negligible.  However, this does not mean that "sputtering" does not occur!  The old technique of PDMS (Torgerson et al. 1974) (plasma desorption mass spectrometry) which is now called "MeV SIMS" (secondary ion mass spectrometry). Recent work (Jones et al. 2011) has demonstrated that the damage cross-section of a sample of leucine (an essential amino acid) deposited on silicon caused by a 10 MeV $^{16}O^{4+}$ is comparable to that of conventional SIMS using keV $Ar^+$ ions – $1.3x10^{14}/cm^2$.

This type of effect was previously demonstrated in a different way by (Gomez-Morilla et al. 2005; Grime et al. 2005), who demonstrated that PTFE could be micromachined by an external proton beam in the presence of oxygen,  although in fact it is not necessary for there to be a high pressure of O.  The same effect can be seen when bombarding PTFE in a (rather poor) vacuum (~$5.10^{-6}$mbar) as can be seen in Figure 19.  This sample was created while using PTFE as a "standard F sample" for a machine energy calibration for a PIGE experiment (Kazor et al. 1994).

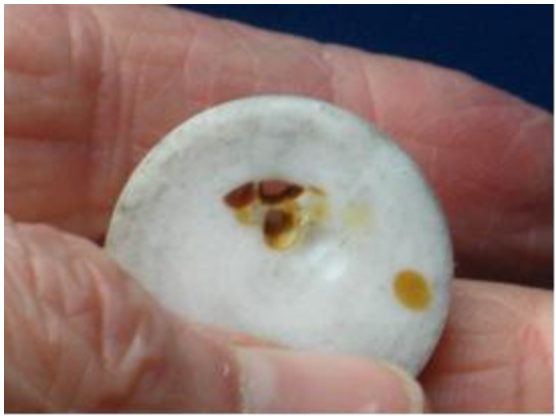

***Figure 19:*** **Enhanced electronic sputtering of insulators.** PTFE disc, 5mm thick.  Irradiated with ~1MeV protons in ~1993-4 to do machine calibration using the resonant F(p,γ) reaction at 990 keV for work reported in (Kazor et al. 1994)

### Hydrogen loss from ion irradiation.

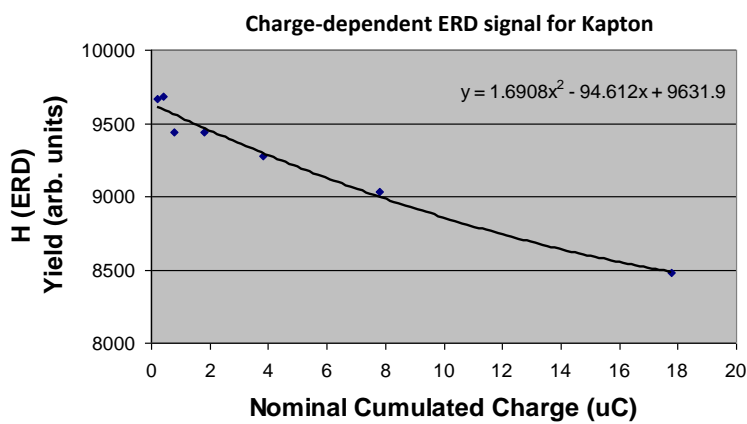

**Charge-dependent ERD signal for Kapton**

$y = 1.6908x^2 - 94.612x + 9631.9$

***Figure 20:*** **Hydrogen loss during ERD of Kapton.** Measurements 1st December 2009, 1433 keV $^4He^+$, glancing incidence 75°, ~15 nA.  Spot size ~4*1 mm, hence flux about 100 nA/cm$^2$ or about 6.10$^{11}$ions/cm$^2$/sec.  Kapton is thick (~0.1mm).

Elastic recoil detection (ERD) using He is a standard method for H profiling. Figure 20 shows measurements made for an ERD analysis during the calibration of the elastic recoil detector solid angle using a standard sample of Kapton ($C_{22}H_{10}O_5N_2$) – well-known as a material resistant to ion beam damage. In the conditions used we had 10% H loss for a fluence of 8 µC into the spot, or about $3.10^{14}$ions/cm$^2$.  The quadratic behaviour is a consequence of H loss from deeper into the sample, where there is a diffusion time to reach the surface.

There is a report (Watamori 2006) that the near-surface hydrogen content of He-implanted thick Kapton (0.2 mm) is almost entirely driven off by annealing up to 300°C.

Polystyrene (PS: $C_8H_8$) shows more pronounced material loss under radiation than Kapton. An early, thorough and heavily cited paper (Abel et al. 1995) explores some features of this process, including the differences between irradiation with H, D, He, and C beams and using "multiscalar" data collection which directly gives the H loss as a function of time. Of course, one thing that is going on is cross-linking, just as for an electron beam. This has been carefully explored (Licciardello et al. 1990). A comparable case is H loss from a-C:H (hydrogenated amorphous carbon) (Adel et al. 1989), and for both a cross-linking process and H-loss from a-C:H one theoretically expects H yield to fall to zero.

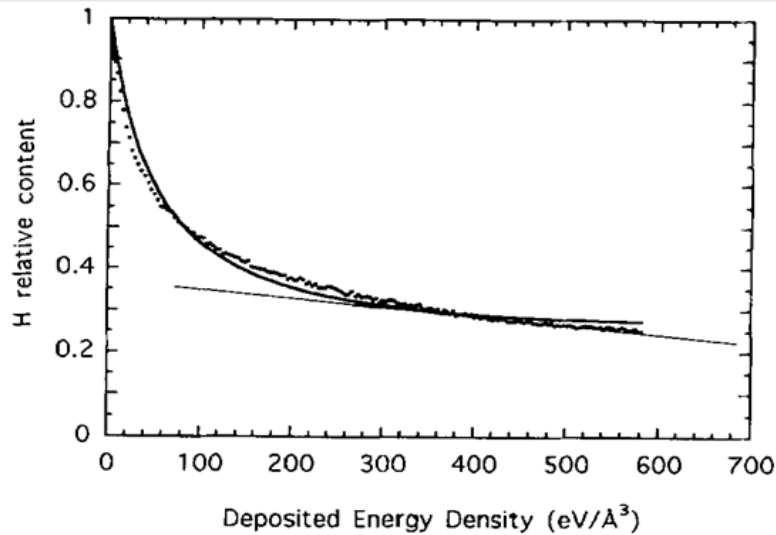

**Figure 21:** shows that for polystyrene (and, in fact, polymers in general) this is not the case. Evolution of H continues to the highest fluences used. This is because the models for cross-linking and for H-loss from a-C:H have a simplified assumption that evolution of H ceases at a certain minimum H-content.

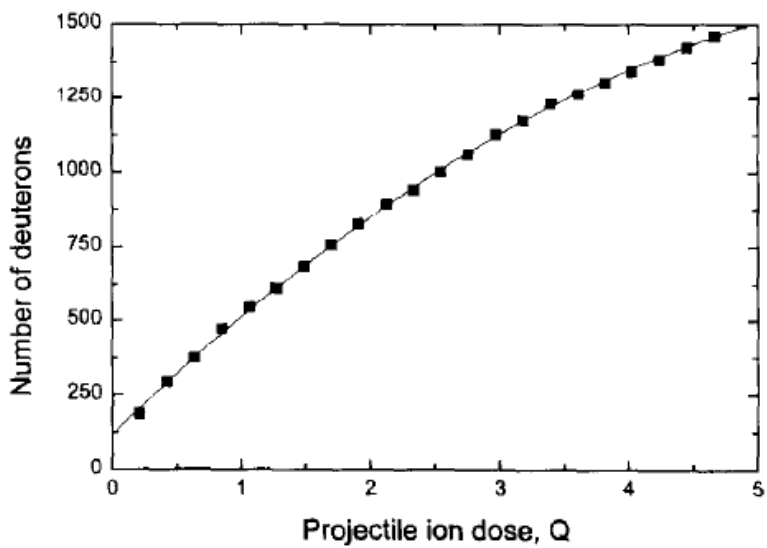

*Figure 22:* **Cumulated D signal from polystyrene for 6 MeV Ne-ERD.** Q is in units of 2. 12. 1014Ne/cm². The line is a quadratic fit to the data showing D loss from displacement events. Reproduced from Fig. 6 of (Ermer et al. 1998)

The maximum analysis fluence limit that radiation damage imposes on the HI-ERD (heavy-ion elastic recoil detection) of d-PS using 4 MeV N & Ne and 6 MeV Ne & Ar beams was investigated by the Freiburg group (Ermer et al. 1998). They base their calculations on the observed thinning of carbon stripper foils under heavy ion bombardment (Dollinger and Maier-Komor 1989), and find experimental support for damage being due simply to displacement events. Figure 22 shows the coincidence of theory and experiment for D loss under the beam. They emphasise what other HI-ERD groups have also shown, that heavier beams are more effective for analysis since the cross-section goes with $Z^2$ but the damage goes much more slowly. Heavy beams cause more damage per incident ion, but give *much* more signal per ion.

### Damage of exit windows and energy degraders

Similar types of measurements were made by (Delto et al. 2002) on PS, but also for the first time on PET (polyethylene teraphthalate: $C_{10}H_8O_4$), which is a more radiation sensitive than PS. (A previous report (Namavar and Budnick 1986) used only RBS, and measurements at low temperatures which showed no discernable compositional change could mean only that the diffusion out of the sample of species evolved under the beam was inhibited.) Delto *et al* used thin transmission films with ERD in coincidence with the forward scatters, which enabled the use of much larger detectors without loss of depth resolution. Incidentally, they suggest that the PET stopping powers are 30% larger than SRIM2000 values. This is unlikely to be a pure Bragg's rule deviation (such deviations have not been observed larger than about 20%) but may very well be due to poor values for the light elements.

Kapton has been investigated as a vacuum window (Matsuyama et al. 1999) at the Tohoku vertical external beam facility. They demonstrated that it could withstand the beam for longer when moderate beam currents were used, where the criterion was whether the windows would break. On the other hand, in their work thinner foils were not so robust, presumably because their heat dissipation was reduced, and of course a unit change in thickness has a larger effect on the fracture strength for a thinner foil. So, using a 3 MeV proton beam for example, a 12.5 μm foil tolerated over 1.5 mC total charge with beam current of 500 nA into a 5.2 mm² spot (about $2.10^{17}H/cm^2$), where a 7.5 μm foil only tolerated just over 0.5 mC total charge with beam current of 1 mA into a 4.2 mm² spot (about $7.10^{16}H/cm^2$).

*Upilex* ($C_{16}H_4O_4N_2$) is a polyimide polymer significantly more radiation resistant than Kapton. It has been investigated as an energy degrader for the Florence external beam facility (Fedi et al. 2002). Of interest in this work was the variation of energy loss of the film due to its thinning under the beam, rather than its fracture strength. Figure 23 summarises some of the results. The initial stage obvious in this Figure where the thickness appears to increase is probably due to material deposition on the sample from the "vacuum" system.

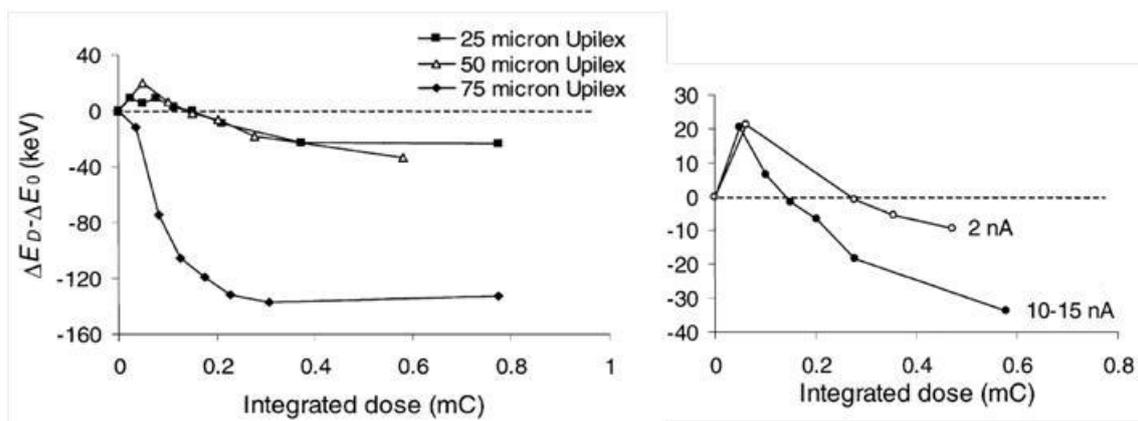

***Figure 23:*** Thinning of Upilex 3 MeV [1]H (0.21 mm² spot). **Left:** Energy loss for various thicknesses of Upilex up to 0.8 mC collected charge for beam currents 10-15 nA ($2.4.10^{18}H/cm^2$). **Right:** Energy loss for various thicknesses of Upilex up to 0.6 mC collected charge for high and low beam currents ($1.8.10^{18}H/cm^2$). Reproduced from figs. 1&2 of (Fedi et al. 2002).

### Polymer damage suppressed by cooling

The Eindhoven group (de Jong et al. 2000) were interested in analysis of polymers used in electronics. Light emitting diodes (LEDs) can be made with significantly longer lifetimes using an organic electrode PEDOT:PSS (*poly-(3,4-ethylenedioxythyophene):poly-styrenesulfonate*) (S, C, H, O) = (6%, 43%, 35%,17%) and an electroluminescent polymer PPV: *poly-(phenylenevynilene)* (C, H, O) = (39%, 57%, 4%). The stability of the PEDOT:PSS/PPV interface, characterised by the mobility of S, is the critical parameter in this device structure. On the face of it, RBS seems to be the ideal technique for this problem. However, the polymers are not sufficiently stable under the ion beam for meaningful analysis to be achievable. So low temperature RBS was carried out using a 2-stage refrigerator capable of 2 W at 10 K. At these temperatures the gas evolved by the ion beam is frozen in place and does not disturb the analysis. Also, extra damage from the high temperatures caused by the energy deposition is eliminated.

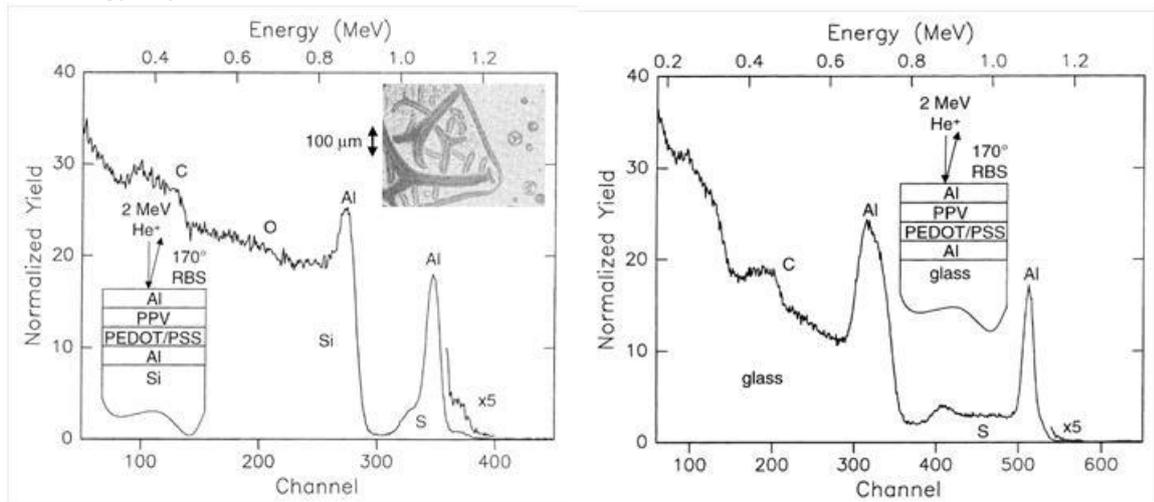

*Figure 24:* **2 MeV He-RBS of polymer LED structures with cooling.** The structure uses Al contacts with an organic electrode (PEDOT:PSS) and the electroluminescent polymer PPV. **Left:** (room temperature RBS) the inset shows the blistering of the Al cap due to bubble formation, and mixing of the structure is characterised by surface S. **Right:** (samples at ~ 50K). Surface S is no longer seen, and all blistering is suppressed. Reproduced from figs. 2&3of (de Jong et al. 2000).

Figure 24 contrasts hot and cold RBS on this system. In the hot system S is easily visible in the PPV film in the as-grown material, showing that degradation of the structure under the ion beam is mixing the interface. But cold RBS gives the correct structure with no change during the analysis. That the cold stage is merely freezing out damage is demonstrated by the blistering that appears as soon as the samples are warmed up. These authors also wished to measure Cl-labelled polyacrylate samples, a material relevant to liquid-crystal displays. But in such samples hot 2 MeV He-RBS/ERD results in massive Cl loss as well as the usual H loss. Cold analysis reduces the H loss by a factor 5 but the Cl by a factor 20.

### Polymer irradiation: direct measurement of volatiles

We have concentrated so far on the evolution of H during irradiation, but of course, many other species are also evolved in these highly energetic processes. (Murphy et al. 2004) have described a series of experiments involving spectrometry of the residual vacuum (using a spectrometer designed for gas chromatography, GC-MS) during irradiation of a variety of polymers. The volume monitored by the GC-MS is separated from the accelerator by a 10 μm Havar window: this degrades the beam energy from 7.5 MeV to an average of 4.25 MeV. Typically 3 cm$^2$ was irradiated by 12.5 nA. (Havar is a high-strength non-magnetic alloy, mostly Co, Cr, Ni; foils of thickness 4 μm also available.)

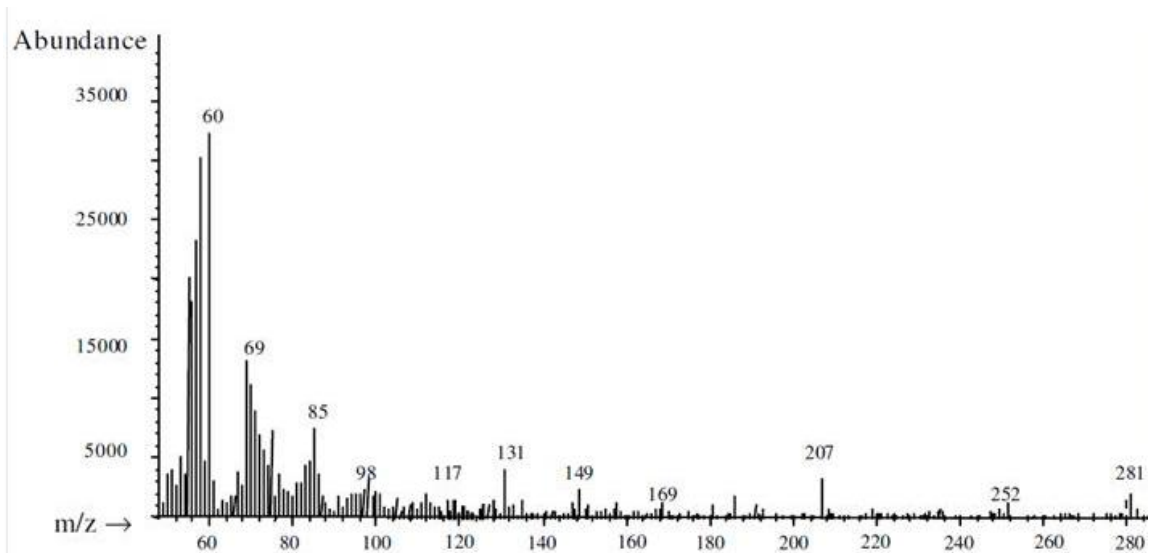

**Figure 25:** GC-MS during He-irradiation of EVA/PVA polymer
Mass spectrum obtained during dose of 1.25 MGy ($55.10^{17}$He/cm²). A similar-looking spectrum is obtained just from the vapour pressure of the polymer, but an order of magnitude lower intensity with all the major peaks in different places. Reproduced from fig.6 of (Murphy et al. 2004)

Three polymers were tested: EVA/PVA (*ethylenevinylacetate/polyvinylalcohol*, with 5 wt% PVA) which is known to be highly unstable under $\gamma$ irradiation; poly(siloxane), known to be very stable; and polyurethane (consisting of a *polyester polyol*, a *diphenylmethane diisocyanate* and an amorphous fused silica filler) with an intermediate stability. It is noticeable that the main species evolved are CO and $CO_2$ (H was not observed with this GC-MS), but many other species are also observed, including ones with relatively high molecular weight (see Figure 25).

The polyurethane was also tested by pyrolysis, and the degradation under the ion beam had similarities with pyrolysis at ~500°C. The effective equilibrium temperature rise expected for the beam power used was <20°C, so it is clear that non-equilibrium (radiation-enhanced) damage effects are being observed.

### Damage to a pollen grain by a proton microbeam

(Watt et al. 1988) describe an experiment where pollen grains of *Impatiens sultanii* were bombarded with a 4 MeV proton microbeam using the Oxford scanning proton microprobe at fluences up to $6.10^{20}$/cm² with some obvious damage as judged by analytical results, and SEM subsequent to the experiment showing some obvious morphological damage (see Figure 26 which shows the effect of a stationary beam and a proton fluence of $3.10^{20}$/cm²). At such high fluences it would indeed be incredible to see no damage!

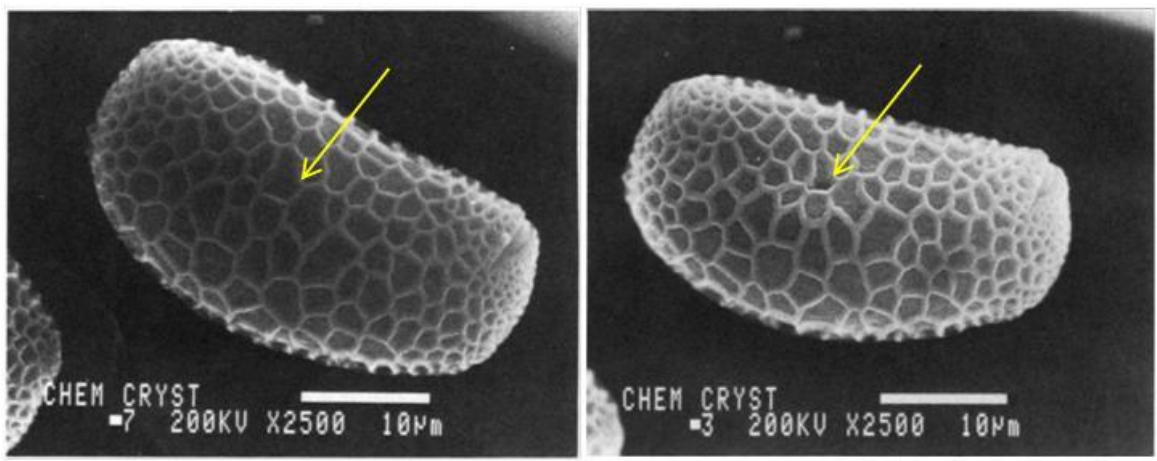

*Figure 26:* **Scanning electron micrographs of gain of** *Impatiens sultanii.* **Left:** Unimplanted; **Right:** Irradiated with 0.5 µC of 4 MeV protons focussed to 1 µm, stationary on sample for about 2h (70 pA). This left an apparent hole (arrowed) with some sample distortion. Reproduced from (Watt et al. 1988).

Detailed analysis of this sample showed that "charring" was visible in the optical microscope (but not the SEM) for proton fluences as low as $3.10^{18}/cm^2$. This was interpreted as a rapid surface carbonisation due to a very rapid H loss (in the first few seconds of exposure). Such an H loss is observed in semiconductor manufacturing when implanting photoresist (see above: *Introduction*), and was also observed on proton microprobe irradiation of a wheat seed (Mazzolini et al. 1981), with similar effects on organic materials in the electron microprobe (Hall and Gupta 1974).

The analytical signal from the (small!) irradiated area of the pollen grain was quite erratic in this experiment, much more than expected from counting statistics. This was probably due to physical distortion of the sample due to the irradiation, exposing slightly different areas of the sample to the beam. The authors conclude that due to this effect accurate analysis using nanobeams (~100 nm) will probably be excluded in this type of sample.

### *Cultural Heritage – Paper*

We here summarise a recent review by Grime (Grime 2011). Paper and other manuscript substrate materials present a significant challenge when subjected to IBA. The materials are often fragile and whitened to enhance the contrast of the ink, which makes the objects particularly sensitive to discolouration. The advice of a leading laboratory (Florence) is instructive for us: from judgements in 1995 (Giuntini et al. 1995) and 1996 (Lucarelli and Mando 1996) that beam currents of, respectively, 11 nA/mm² and 3nA/mm² were not damaging to precious manuscripts, ten years of experience yielded more caution: even ¼ nA/mm² was observably too much in some cases (Grassi et al. 2007). To exclude ion beam damage from paper, not only must the analysis beam intensity be strictly limited but also the analysis must be done in air with forced cooling using a helium jet (helium is a very good conductor). And it should be noted that even though IBA may not leave visible damage, it may still may leave damage that is visible under UV (McKay 1995).

(Cheng et al. 1998) in an interesting and informative paper, demonstrated that the flux density was not critical provided it was below a certain value, but that the paper strength became degraded with total fluence over a certain value. The critical values depend on the circumstances of course, but for 7 mg/cm² copy paper the upper limit of the beam flux is 1.2 and 0.8 µC/cm² for proton energies of respectively 3.3 and 2.3 MeV (neither of which is stopped by this paper): that is, the paper can withstand about 3 J/cm$^2$ of energy deposition (in this case about 0.4 Gy) without damage. In this work the critical flux was < 0.15 nA/mm$^2$, consistent with the later (2007) reports from Florence.

## 2.8 Maximum Data Extraction

### Overview
Data analysis is the process of extracting information from the data collected. This is relevant to ion beam induced damage produced during the experiment in two fundamental ways. The first is how to meaningfully analyse data obtained when Type I damage is incurred. The second is when the experimentalist, aware that the sample might change under the beam, consciously decides to collect low statistics data, from which nevertheless a maximum amount of information should be extracted.

That is, on the one had beam damage can lead to data that do not represent the sample as it was before the experiment; and on the other hand, preventing beam damage from occurring can lead to poor quality data. In this section we discuss strategies to cope with these issues.

Chapters 14 ('Data analysis for IBA') and 15 ('Pitfalls in IBA') of the 2$^{nd}$ edition of the Handbook of Modern Ion Beam Analysis (Wang and Nastasti 2010), provide numerous references for further reading. Novice ion beam data analysts are urged to read those chapters carefully. But first read this short introduction to the subject.

### Damaged data
We first consider Type I damage, illustrated by an example. Consider Fig. 27 that shows the RBS and $^4$He-ERDA data of an InN sample, irradiated with a fluence of 13.1×10$^{15}$ $^4$He/cm$^2$ (from Lorenz et al. 2011).

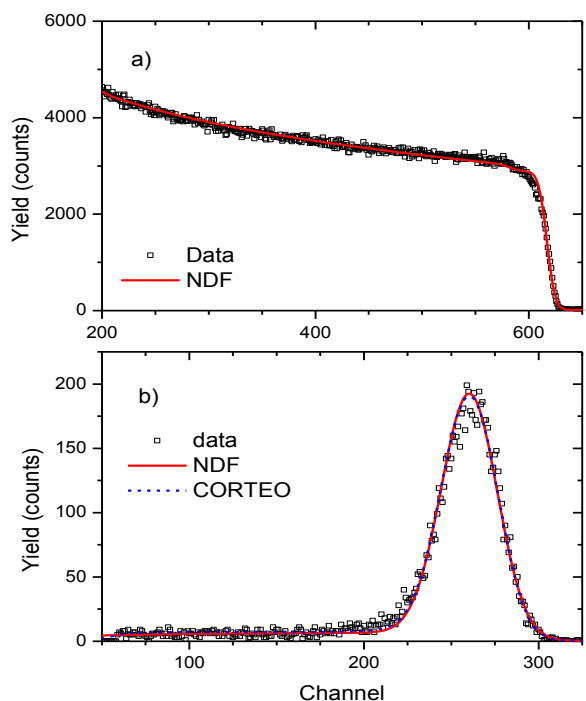

The surface hydrogen areal density and the hydrogen concentration in the sub-surface region can be easily determined from this spectrum, which has no indication of beam induced damage. However, in this case the analysts suspected hydrogen loss during the experiment, so they collected spectra, from the same spot, as a function of the $^4$He beam fluence. The results are shown in Figure 28, and prove a very strong hydrogen loss, that can be quantified with this method. We note, however, that if one is not interested in hydrogen, and because its effect on the RBS data is small, the results for the other elements are practically unaffected by the hydrogen loss, and would be correct.

In extreme cases, sputtering can lead to a complete loss of the sample, if it was a thin film on a substrate. Often, it will lead to only a changing thickness of the surface layer, from the initial value down a smaller thickness. If there is no preferential sputtering of some element, this can be simulated in modern data analysis codes by inputting a thickness distribution for the affected layer (Mayer 2002).

**Figure 27** RBS and $^4$He-ERDA data of an InN sample

However, we note that the effect on the data is very similar to that of surface roughness, and if the data analyst is not aware that sputtering occurred, the results given back to the sample owner could be in terms of an average thickness plus a roughness value - and this would be misleading, since the sample could have been initially perfectly flat, and its thickness was for certain larger than the final value quoted. To analyse the data correctly, the data analyst needs to know that sputtering occurred, and probably also needs to know how much.

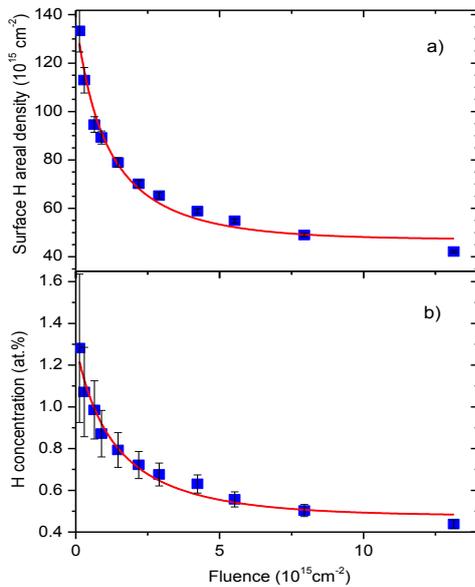

**Figure 28:** hydrogen loss as a function of beam dose during $^4$He-ERDA of an InN sample

The analysing beam can lead to ion beam mixing. If the data analyst is not aware of this, the analysis may incorrectly consider that the mixing was pre-existent, or falsely assign the broadened data to roughness.

It is clear that there is no general method to deal with damaged data. The most important thing is to realise that the data may not represent the sample as it was before the measurement, otherwise a perfect fit can mean that a useless analysis is taken as correct. That said, it is often impossible to tell from a given spectrum that damage actually occurred. So, either there is pre-existing knowledge about the system, or the knowledge is obtained via a different technique, or data are collected as a function of analysing beam fluence (ideally in list mode). This is only usual in heavy ion ERDA or where a scanning microbeam is used. In other techniques it is only done when there is an a priori expectation of beam damage.

### Low statistics data : minimising Type II damage by minimimising the incident beam fluence

One possible preventive action to avoid beam damage is to collect low statistics data, minimising the beam fluence. The question is whether low statistics data are useful at all. The answer is a conditional yes depending on the problem. Consider the top panel of Figure 29 from (Barradas et al. 1999). It is a constructed depth profile of an imaginary CoFe silicide multilayer on Si, with a surface SiO$_2$ layer.

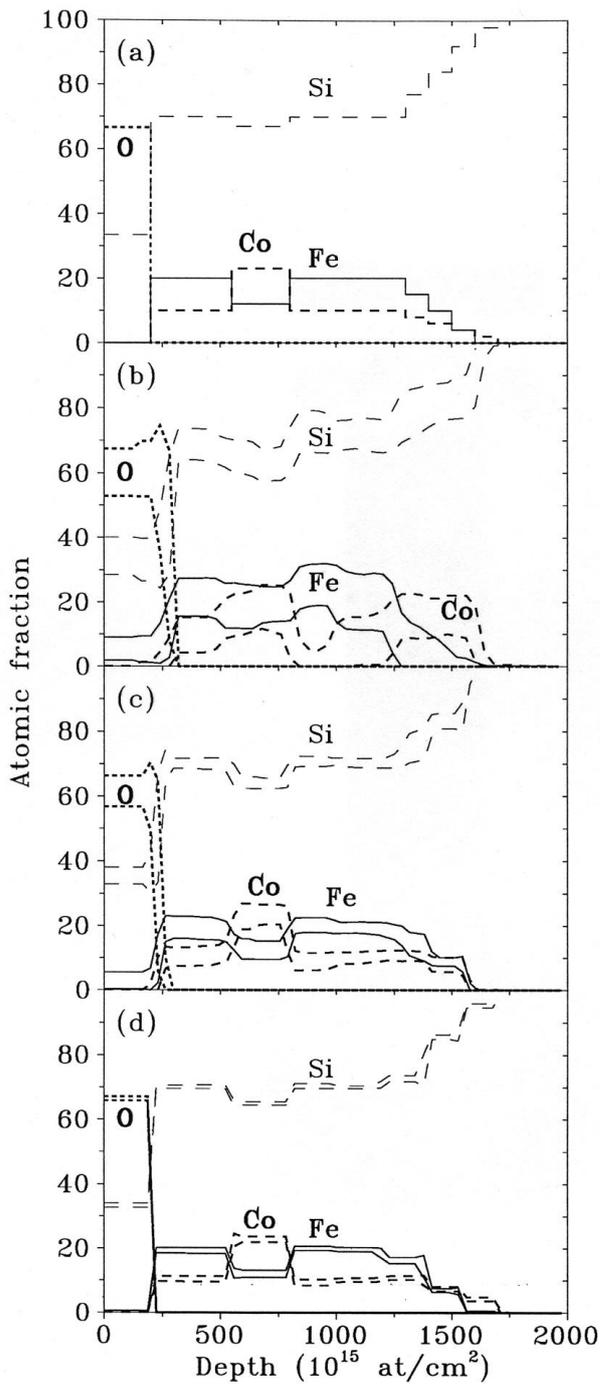

**Figure 29:** (a) Theoritical depth profile used to generate the test RBS spectra analysed, with (b) high, (c) medium and (d) low level of noise. Confidence limits (± 1 standard deviation) of the posterior probability distribution obtained with MCMC for the different elements. Oxygen was only allowed to exist at the surface of the sample. Co and Fe were restricted to depths below $2 \times 10^{18}$ at/cm².

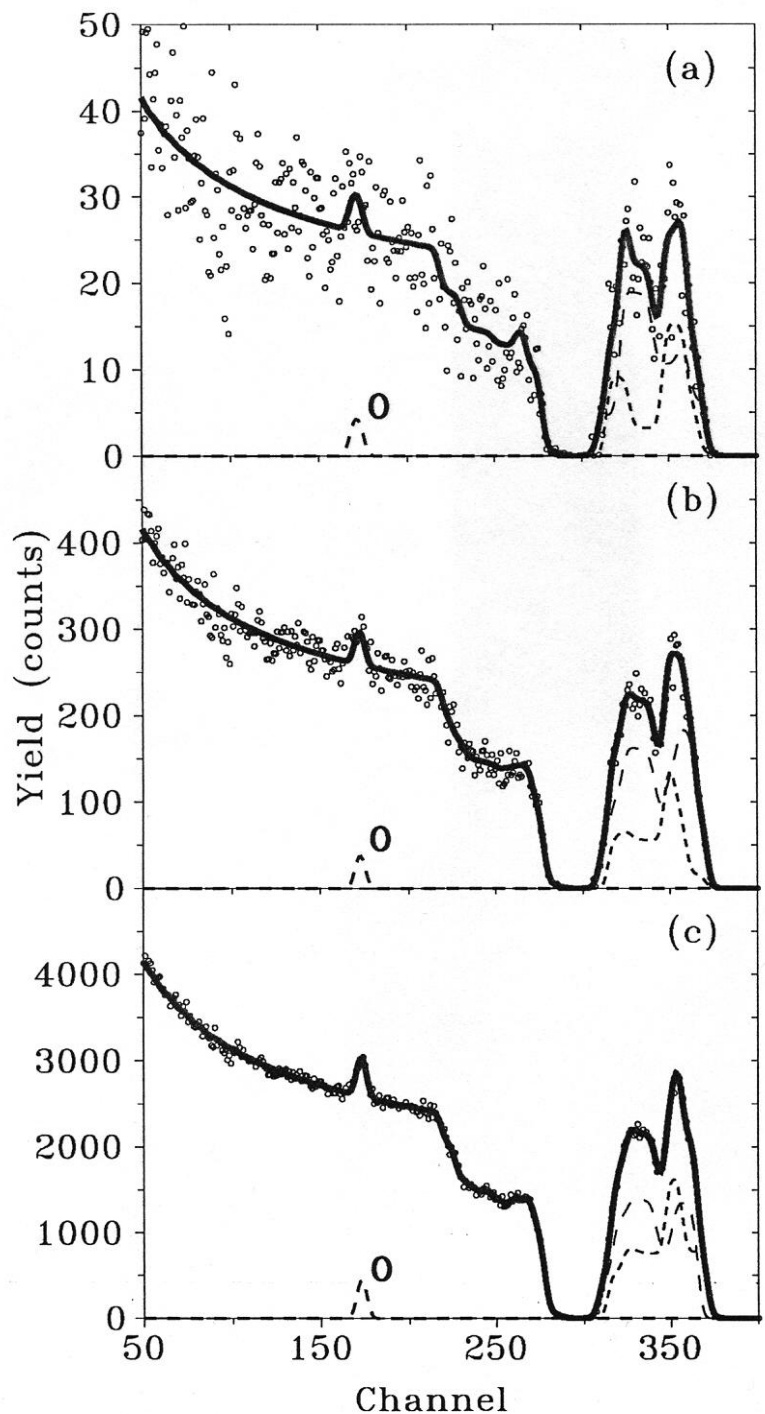

**Figure 30:** Theoritical test RBS spectra (points) and simulated annealing fit (solid line) obtained for the structure shown in Fig. 1, for (a) high, (b) medium and (c) low level of noise, corresponding to beam fluencies values of 0.1, 1 and 10 µC, respectively. The Partial fitted spectra due to the Fe (long dash), Co (small dash) and surface O are also shown as dashed lines.

The corresponding theoretical spectrum was calculated, but adding Poisson noise corresponding to different statistics, in order to simulate something similar to experimental spectra. Spectra with low, fair, and good statistics were generated. They are shown in Figure 30 (they are not the solid lines, they are the points that look like real data). Then, an automated fit was done to each one of them using the code NDF. The fits are the solid lines in figure 30, and the partial fitted spectra of Co and Fe are also shown.

Not stopping there, a Bayesian inference analysis was made with NDF in order to calculate the limits of confidence for the depth profile derived from each of the spectra. The results are shown in Figure 29. It is clear that, even in the low statistics spectra, the correct depth profile is retrieved, but with rather large uncertainties as expected.

So, in this particular case, if the information required is whether this is a CoFe silicide sample with several layers with changing Fe and Co concentration, and what are the approximate thicknesses and concentrations of each layer, then the low statistics spectrum would suffice. In fact, it would be completely unnecessary to collect more statistics. If, on the other hand, more details were required, such as sharpness of the layers, or a good accuracy in their stoichiometry, then clearly more statistics would be needed.

### Multiple spectra

One way of collecting better statistics data while minimising the beam fluence, is to use large solid angle detectors. These, however, usually suffer from low resolution. One alternative is to use multiple detectors located at different positions. The statistics of each one can be low, but the different angles mean that the data are not equivalent, that is, they carry complementary information, and thus lead to a reduction in the ambiguity that is inherent to backscattering. Furthermore, when performing BS or ERDA experiments, the characteristic X-rays are usually discarded, even when they would be extremely useful: in fact, the combination BS/PIXE is extremely powerful due their complementarity. The BS/ERDA and PIXE data should always be collected simultaneously, and a PIXE detector should be installed in every backscattering setup.

The question arises of what to do with so much data. The code NDF is designed to cope with that, by analysing all the spectra simultaneously with the same depth profile. All the information present in each spectrum is extracted, and combined into a best global solution. This includes RBS, EBS, ERDA, non-resonant NRA, resonant NRA, and PIXE with both H and He (and Li etc) beams. This is an extremely powerful enabling tool.

Multiple spectra are also a way to check for inconsistencies due to ion beam damage. For instance, the effect of roughness on the data depends on the detection angles, and it may be possible to distinguish ion beam mixing from roughness when detectors at both near-normal and grazing detection are used. Also, an elemental surface barrier moving due to sputtering may be easier to observe in some geometries than in others. In general, enforcing consistency between different measurements is one of the best ways to avoid wrong data analysis.

### Uncertainty budget

IBA techniques are often hailed as inherently quantitative, without need for external standards. This may be true, but more often than not results are presented without the associated uncertainties. And when "errors" are given, they are almost always the statistical uncertainty only, which is often a small part of the total uncertainty. This is not really acceptable! Confronted with low statistics or with damaged data, the analyst must make the extra effort to calculate what the uncertainty actually is, including all its sources.

The best way of doing this is to make an uncertainty budget. This is a formal systematic evaluation of the uncertainty of an experiment, considering all factors involved (Sjoland et al. 2000). Chapter 15 of (Wang and Nastasi 2010) presents one complete example, that includes two detectors. More generally, one can include multiple spectra in the uncertainty budget. The sources of error include the counting statistics, the uncertainties in the experimental parameters, uncertainties in the algorithms used to analyse the data (such as pileup correction and numerical accuracy of the codes used), and, very importantly, uncertainties in the basic databases used in the analysis. That is, the cross section and the stopping power, which are often taken for granted, but often their uncertainty dominates the final uncertainty in the results quoted. We strongly advise everyone to actually do an uncertainty budget at least once, to understand exactly what it is, and to always do it when the accuracy of results is critical. We note however that the uncertainty budget only allows us to calculate the uncertainty in specific parameters (such as the concentration of a given element in a given layer), and not the uncertainty in a depth profile.

One alternative is to use NDF to calculate the uncertainty. NDF implements a Bayesian inference routine to calculate uncertainties, in both specific parameters and in the depth profile as whole. Figure 29 shows one example. During the Bayesian inference analysis (which can be very time consuming), the experimental parameters are allowed to change freely within their experimental uncertainties. At the same time, the depth profile also changes. In the end, not one but very many depth profiles are obtained, all of them consistent with the data, within statistical errors and within the uncertainty of the experimental parameters. NDF outputs the average and standard deviation of all the depth profiles that are consistent with the data, that is to say, the uncertainty in the depth profile is the final result of the analysis. One weakness of this procedure is that the uncertainties in the cross section and stopping powers are not included. Another weakness is that it relies on complex calculations that are a black box, so it is not transparent. Finally, the values obtained depend on how good the models in NDF represent the experiment at hand. While these are often excellent, in some cases they are not, and in those cases the uncertainties calculated can be wrong.

### *Overinterpretation and underinterpretation*
Over-interpretation and under-interpretation are the two capital sins of the data analyst, and great care must be taken to ensure that these pitfalls are avoided.

Over-interpretation happens when the analyst unjustifiably imposes a given model on the data, when other models would also lead to an equivalent or better solution. Consider a multilayer sample nominally Si/Re buffer/ (Co 2 nm/Re 0.5 nm)×15. The sample owner wishes to know whether this nominal structure is real, or whether the Co and Re are mixed. The spectrum collected at near normal incidence (45º with the sample surface) is shown in Figure 31, together with a simulation made assuming the nominal sample structure (with minor changes in the layer thickness). A report that states "the data are consistent with the nominal structure" is not justifiable: it is hopeless over-interpretation. Indeed, such a report might be understood as corroborating the multilayer structure, of which there is no sign whatsoever in the spectrum. It is perfectly possible to fit the data with a simple Si/Re buffer/CoRe structure. The report should state all of this: that the data are consistent both with a nominal structure and with a uniform CoRe film, and that more data are needed with better depth resolution.

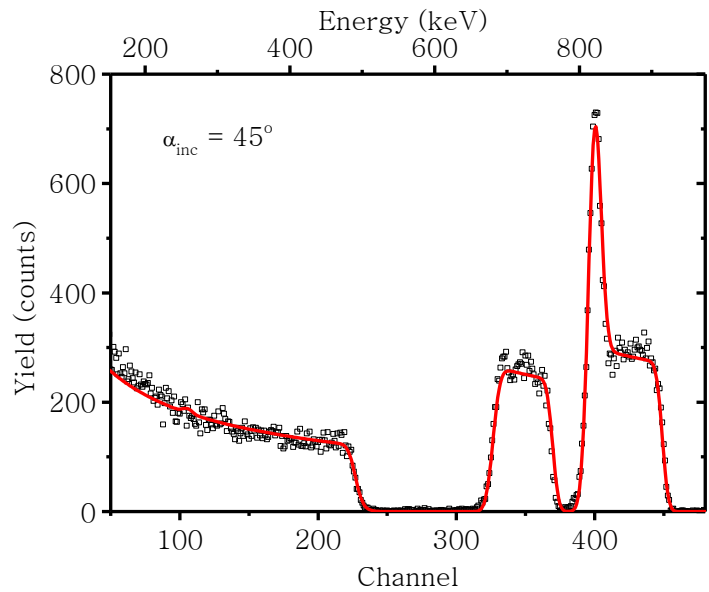

**Figure 31:** RBS spectrum of multilayer sample nominally Si/Re buffer/ (Co 2 nm/Re 0.5 nm)×15 (points) and simulation assuming ideal multilayer structure (line).

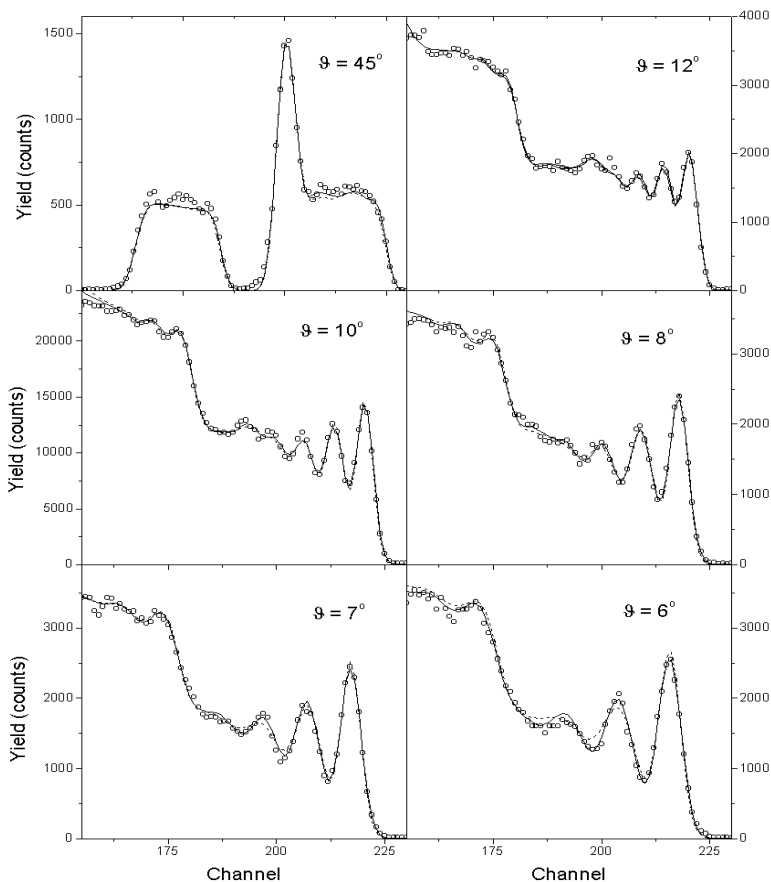

**Figure 32:** Grazing incidence RBS spectra from the sample of figure 31 clearly showing multilayer structure.

Under-interpretation happens when the analyst does not extract all information that could be extracted, often due to lack of knowledge or lack of proper code. Figure 32 shows grazing incidence data from the same sample as shown in Figure 31. These data prove that indeed it is a multilayer. A good report states that the data are consistent with the nominal structure and gives the average and the standard deviation of the thickness of the Co and Re layers, but it does not stop there. In fact, the data carry information about the presence of roughness, and are enough to actually quantify it. However, a code that can do it, and has the proper model of roughness, is required. A good report, avoiding the sin of under-interpretation, makes use of such a code and quantifies the roughness.

# 3.    Recommendations for damage mitigation

Damage can never be eliminated : the name of the analysis game is optimisation of the information to damage ratio. In a successful analysis enough information will be extracted from the sample in order to answer the scientific question posed, whilst incurring insufficient damage to compromise either the analysis itself, or any future requirements that may be made of the sample (e.g. further examination of physical or chemical properties, long term archivage of environmental samples, integrity of valuable museum or forensic samples and so on). Common sense and a few simple experimental optimisations aimed at minimising the beam dose will go a long way to reducing the damage incurred during an analysis. Reducing the beam dose requires firstly determining the minimum analytical signal that will supply the required analytical answer, and then optimising the experiment so that this smallest useful analytical signal is generated with the lowest beam fluence (and flux, where flux dependant effects such as beam heating are significant). One further strategy that is useful is to minimise energy accumulation in the sample, for example by providing good conduction pathways to remove heat from the sample.

## 1) Know your sample, and it's future

Damage induced during IBA is highly sample dependant, and the first step to successfully avoiding unnecessary sample damage is for the analyst to know as much as possible about the sample, in order to judge the likelihood of damage being a problem. This means knowing the physico-chemical nature of the sample (is it a conductor? Is it particularly heat sensitive? Does it contain chemical bonds likely to be disturbed by beam ionisation? Could radiation-enhanced diffusion or ballistic mixing be a problem? ...). In some cases this may not be known, in which case preliminary experiments with a sacrificial sample may be necessary. If it turns out that damage could be a significant problem, then mitigation measures are necessary.

Type II damage may be quite acceptable, for example when a sample is prepared uniquely for the IBA analysis. The owner abandons the sample once the IBA analysis is performed and the sample has no future. At the other extreme, the sample may be the result of a long or difficult preparation process, be particularly valuable financially, culturally or scientifically, or be subject to further characterisations. In such cases Type II damage may be a major problem (modification of electrical or optical properties, visible or latent damage on art works, compromising forensic samples, etc).

## 2) Determine the analytical requirement

The main point here is to know the level of uncertainty in the measurement that must be attained in order to reply adequately to the analytical question that is posed. For example if the intention is to determine the difference in the oxygen content of two thin films with 0.2% precision high statistics will be necessary (around a million analytical counts for each oxygen determination) – however if the intention is to determine oxygen contamination with 5% uncertainty then a few hundred analytical counts are all that is required.

One aspect of determining the analytical requirement that is often neglected is the consideration of just what information is contained in a spectrum. If the analytical signal consists of a single isolated peak and systematic errors are kept sufficiently low then the uncertainty may be dominated by the counting statistics. In more complicated spectra where the link between the spectrum and the analytical information is more complex, more sophisticated methods are required to determine what information is, and what information is not, contained in a spectrum. A fully analytical approach to answer this question is hardly practical for every analysis, however much can be learned from simulations and from informed use of automated data fitting routines (such as WINDF, which estimates uncertainties in the analytical information derived from multiple spectra) on data simulated for a given analysis with different counting statistics.

### 3) Minimise the beam dose and dose density

If damage is judged to be a likely problem, the first experimental optimisation is to minimise the beam dose necessary to obtain spectra of adequate statistics. Obvious measures such as using the maximum detector solid angle, using the largest beam diameter (to spread the required dose over as large an area as possible), and choosing the highest yielding IBA technique where more than one can give the required information should be routinely applied for all IBA analyses.

- Maximising the detector solid angle may require the use of multiple detectors since detector energy resolution and surface area are often inversely related (for example in silicon charged particle detectors, and energy dispersive X-ray detectors). One SPIRIT research activity has centred on development of an array of charged particle detectors closely spaced on a single silicon chip so as to achieve high surface area (and hence solid angle) without compromising either detector resolution or kinematic broadening for RBS and light particle ERDA. On a slightly less sophisticated level, an increasing number of IBA laboratories routinely use 2, 3 or more conventional detectors for RBS, NRA and/or PIXE. Use of multiple gamma detectors is also well established.

- Using a large beam may not always be feasible. For example it may not be reasonable to assume that the sample is homogenous over a large enough area. In light particle ERDA a large beam impact area results in a significant spread of angles subtended by even the smallest detector, and in many installations ion beam optics may not be optimised to produce sufficiently large homogenous beam spots. For homogeneous samples, and for situations where the beam is deliberately focussed (nuclear microprobe) beam or sample scanning can be envisaged. In Medium Energy Ion Scattering with an electrostatic spectrometer, where tight detection geometry (including incident beam size) is crucial to obtain good energy resolution, sample scanning is routinely applied in order to spread the beam dose over as large an area as possible.

- Providing good thermal pathways for heat dissipation can be very effective at minimising temperature rise. Mounting the sample on a significant low temperature reservoir (often the sample holder, which constitutes a large thermal reservoir at ambient temperature, and which may be cooled if necessary) with good thermal contact should be routine. It is worth remembering, also, that in thin transmission samples the incident beam deposits only a portion of its energy. For higher energy beam particles, for which the stopping power is smaller, the energy deposited in the thin sample is actually less than for a lower energy beam. In external beam IBA conductive cooling via the surrounding gas (air, He, possibly supplied via a nozzle and jet to ensure a continuous supply of gas at ambient or reduced temperature) is very effective.

- In cases where Type I beam damage can be observed during the analysis, one approach that has been successfully used is to register the spectra in list mode – where each detection event triggers storage of the event parameters (at a minimum: particle energy, time of detection, and integrated incident beam dose at time of detection) in a list. From such a list, after the end of the measurement, partial spectra may be retrieved for various small increments of dose. The analytical signal of interest may then be plotted as a function incident beam dose and if a smooth or understandable relationship is found between the beam dose and the analytical signal, extrapolation of the analytical signal back to zero beam dose may give the most reliable estimate of the analyte. If list mode data acquisition is not available, an alternative is to register sequential spectra for incremental beam doses.